\documentclass[apj]{emulateapj}
\usepackage{ amsmath, color, natbib}
\newcommand{\kms}{km~s$^{-1}$}
\newcommand{\Myr}{M$_\sun$~yr$^{-1}$}
\newcommand{\Msun}{M$_\sun$}

\begin{document}
\title{A far-IR view of the starburst driven superwind in NGC 2146}

\author{K. Kreckel\altaffilmark{1}, 
L. Armus\altaffilmark{2}, 
B. Groves\altaffilmark{1}, 
M. Lyubenova\altaffilmark{1}, 
T. D\'{i}az-Santos\altaffilmark{2}, 
E. Schinnerer\altaffilmark{1}, 
P. Appleton\altaffilmark{3}, 
K. V. Croxall\altaffilmark{4}, 
D. A. Dale\altaffilmark{5}, 
L. K. Hunt\altaffilmark{6}, 
P. Beir\~{a}o\altaffilmark{7}, 			
A. D. Bolatto\altaffilmark{8, 9},			
D. Calzetti\altaffilmark{10}, 
J. Donovan Meyer\altaffilmark{11}, 		
B. T. Draine\altaffilmark{12},  			
J. Hinz\altaffilmark{13, 14}, 			
R. C. Kennicutt\altaffilmark{15}, 		
S. Meidt\altaffilmark{1},				
E. J. Murphy\altaffilmark{16}, 			
J. D. T. Smith\altaffilmark{17}, 			
F. S. Tabatabaei\altaffilmark{1}, 		
F. Walter\altaffilmark{1}				
} 

\altaffiltext{1}{Max Planck Institut f\"{u}r Astronomie, K\"{o}nigstuhl 17, 69117 Heidelberg, Germany;  kreckel@mpia.de}
\altaffiltext{2}{Spitzer Science Center, California Institute of Technology, MC 314-6, Pasadena, CA 91125, USA}
\altaffiltext{3}{NASA Herschel Science Center, IPAC, California Institute of Technology, Pasadena, CA 91125, USA}
\altaffiltext{4}{Department of Astronomy, The Ohio State University, 140 West 18th Avenue, Columbus, OH 43210, USA}
\altaffiltext{5}{Department of Physics and Astronomy, University of Wyoming, Laramie, WY 82071, USA}
\altaffiltext{6}{INAF-Osservatorio AstroÞsico di Arcetri, Largo E. Fermi 5, I-50125 Firenze, Italy}
\altaffiltext{7}{Observatoire de Paris, 61 avenue de l'Observatoire, Paris F-75014, France}
\altaffiltext{8}{Department of Astronomy and Joint Space-Science Institute, University of Maryland, College Park, MD, USA}
\altaffiltext{9}{visiting Humboldt Fellow, Max-Planck Institute for Astronomy, Heidelberg, Germany}
\altaffiltext{10}{Department of Astronomy, University of Massachusetts, Amherst, MA 01003, USA}
\altaffiltext{11}{National Radio Astronomy Observatory, Charlottesville, VA 22901, USA }
\altaffiltext{12}{Princeton University Observatory, Peyton Hall, Princeton, NJ 08544-1001, USA}
\altaffiltext{13}{Steward Observatory, University of Arizona, Tucson, AZ 85721, USA }
\altaffiltext{14}{MMT Observatory, Tucson, AZ 85721, USA}
\altaffiltext{15}{Institute of Astronomy, University of Cambridge, Madingley Road, Cambridge CB3 0HA, UK}
\altaffiltext{16}{Infrared Processing and Analysis Center, California Institute of Technology, MC 220-6, Pasadena CA, 91125, USA}
\altaffiltext{17}{Department of Physics and Astronomy, University of Toledo, Toledo, OH 43606, USA}
\begin{abstract}

NGC~2146, a nearby  luminous infrared galaxy (LIRG), presents evidence for outflows along the disk minor axis in all gas phases (ionized, neutral atomic and molecular).  We present an analysis of the multi-phase starburst driven superwind in the central 5 kpc as traced in spatially resolved spectral line observations, using far-IR Herschel PACS spectroscopy, to probe the effects on the atomic and ionized gas, and optical integral field spectroscopy to examine the ionized gas through diagnostic line ratios.  We observe an increased $\sim$250~\kms~velocity dispersion in the [OI] 63~$\mu$m, [OIII] 88~$\mu$m, [NII] 122~$\mu$m and [CII] 158~$\mu$m fine-structure lines that is spatially coincident with high excitation gas above and below the disk.  We model this with a slow $\sim$200~\kms~shock and trace the superwind to the edge of our field of view 2.5 kpc above the disk.  We present new SOFIA 37~$\mu$m observations to  explore the warm dust distribution, and detect no clear dust entrainment in the outflow.  The stellar kinematics appear decoupled from the regular disk rotation seen in all gas phases, consistent with a recent merger event disrupting the system.  We consider the role of the superwind in the evolution of NGC~2146 and speculate on the evolutionary future of the system.  Our observations of NGC~2146 in the far-IR allow an unobscured view of the wind, crucial for tracing the superwind to the launching region at the disk center, and provide a local analog for future ALMA observations of outflows in high redshift systems. 

\end{abstract}


\section{Introduction}

Galaxy outflows driven by star formation and active galactic nuclei (AGN) allow redistribution of energy and metals within the interstellar medium \citep{Veilleux2005} and have the potential to strongly affect galaxy evolution through quenching of star formation \citep{Kormendy2009}.  They are a common feature in luminous infrared galaxies (LIRGs) \citep{Rupke2005}, which are observed to increasingly dominate the  star formation rate density at higher redshifts up to z $\sim$2 \citep{LeFloch2005, PerezGonzalez2005}.  

NGC~2146 is one of the closest (17.2 Mpc, \citealt{Tully1988}) infrared luminous (L$_{\rm IR} = 1.2 \times$ 10$^{11}$ L$_\sun$, \citealt{Sanders2003}) galaxies.  With a stellar mass of $2 \times 10^{10}$ M$_\sun$ \citep{Skibba2011} and morphological type SB(s)ab pec \citep{deVaucouleurs1991}, it displays a disturbed optical morphology due to a merger and a bright central bulge, extended irregular spiral arms, and deep dust lane features \citep{Greve2006}.  H~\textsc{i} imaging reveals an extremely extended ($\sim$200 kpc) tail \citep{Fisher1976, Taramopoulos2001}.  NGC~2146 has a well established superwind along the minor axis that has been detected in X-rays and optical emission lines from ionized atomic gas that show evidence for shock excitation \citep{Armus1995, Greve2000} as well as in a molecular gas outflow and superbubbles \citep{Tsai2009}. There is no evidence for AGN activity in the optical \citep{Ho1997} or in mid-IR IRS spectra \citep{Bernard-Salas2009, Petric2011}, though it may host a low-luminosity AGN based on its compact nuclear X-ray emission \citep{Inui2005}.  As it exhibits a relatively high star formation rate of 7.9 M$_\sun$ year$^{-1}$ \citep{Kennicutt2011}, stellar winds and supernovae are thought to drive the superwind, similar to that observed in the starburst galaxy M82 \citep{McCarthy1987, Armus1989, Heckman1990}.  However, as NGC~2146 is both larger and more massive compared to M82, its outflow exhibits a larger geometric scale and slower wind velocities \citep{Greve2000}.  

\begin{figure*}[!ht]
\centering
\includegraphics[width=6.5in]{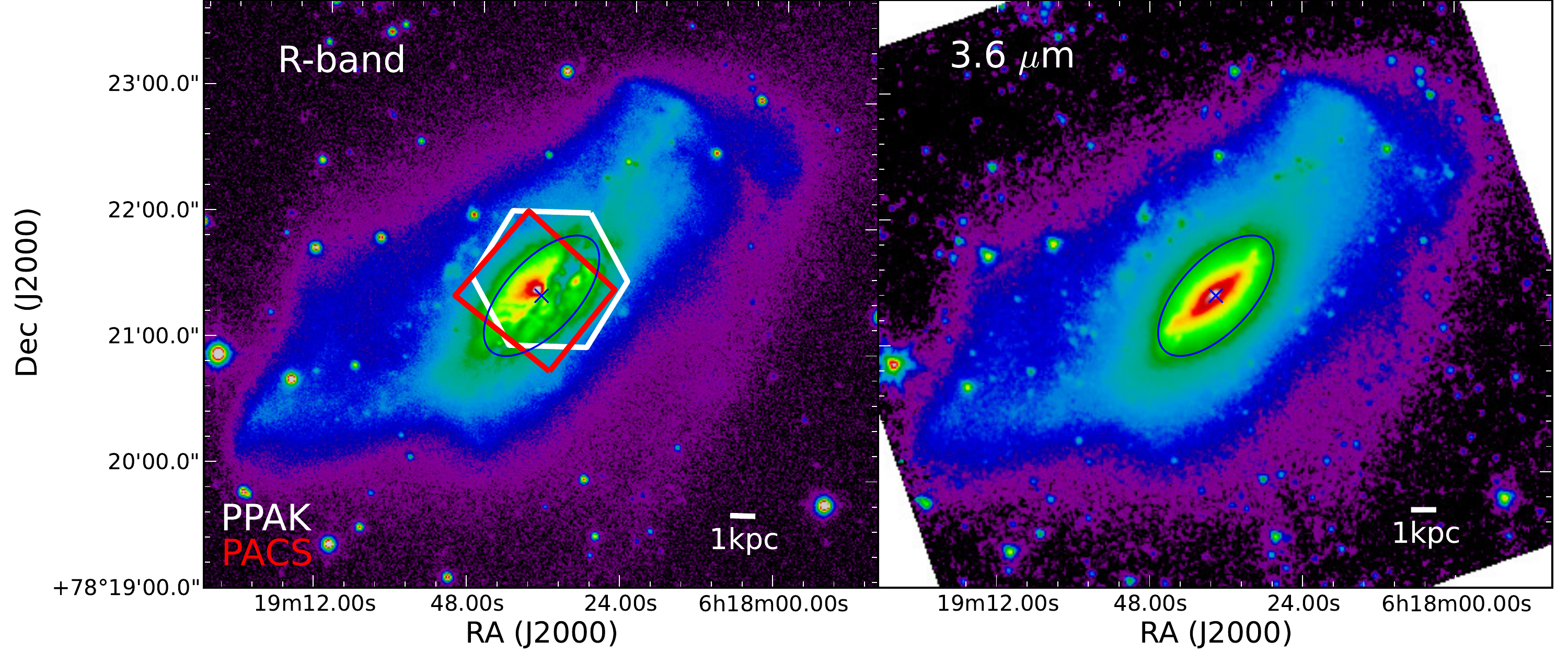}
\caption{R-band (left) and Spitzer/IRAC 3.6~$\mu$m (right) images of NGC~2146.  Both are shown with a log scaling to show low surface brightness features.  Overlaid are the field of view for the PPAK (white) and PACS spectroscopy (red) instruments, as well as an ellipse (blue) in order to provide a common reference for the remaining figures in this paper, with the center as determined from CO kinematic modeling \citep{Tsai2009} marked with an x. The ellipse corresponds to a diameter of 6 kpc at the assumed distance.
\label{fig:fov}}
\end{figure*}

Extensive analysis of the ionized and molecular gas kinematics have resulted in an understanding of the geometry of the system \citep{Greve2000, Tsai2009}.  The disk is highly inclined at 63$^\circ$ \citep{dellaCeca1999}, with the near side of the disk in the south-west,  making the south-west  outflow behind and the north-east outflow in front of the disk.  The outflow follows a conical morphology above and below the disk.  The molecular gas outflow is seen through a breakout on the north-east side and two superbubbles, one to the south-east and another in projection along the major axis \citep{Tsai2009}.  The combination of disk inclination angle and cone opening angle results in an unusual line-of-sight projection of the system, as diagramed in \cite{Greve2000} and \cite{Tsai2009}.  The far side of the cone to the north and the near side of the cone to the south are essentially perpendicular to the line of sight, making their velocities consistent with the systemic velocity.  The other walls of the cone fall, in projection, over the bright central disk component and may not be easily seen.  

Studies of nearby LIRGs and ULIRGs show they are generally interacting systems \citep{Armus1987, Sanders1988, Murphy1996}, and integral field spectroscopic surveys have revealed complex kinematics in multiple gas phases \citep{Arribas2008, Alonso-Herrero2009, Alonso-Herrero2010}.  These kinematics develop along a merger sequence towards more elliptical or lenticular dynamics \citep{Bellocchi2013}. Many show evidence for shock ionization \citep{Armus1989, Monreal-Ibero2010}.  However, optical studies are restricted by dust extinction, particularly along the galaxy disk and in the center where these winds are launched.  Far-IR observations are able to penetrate these dusty regions, providing a direct view of the complete galactic wind.  Such outflows have been observed within ULIRGs by molecular absorption against the nucleus \citep{Fischer2010, Sturm2011, Veilleux2013, Spoon2013}, but the possibility also exists to map outflow regions through the far-IR fine-structure emission lines.

We present here results from Herschel PACS spectroscopy, which reveal conical outflows in the atomic and ionized gas that can be traced back to the central region of the galaxy disk.  We link this to new optical integral field spectroscopy that traces the ionized shock diagnostics over a broad region both above and below the galaxy disk.  We also examine the warm and cold dust distribution through multi-wavelength far-IR imaging.  We present our observations in Section \ref{sec:data}, and our results in Section \ref{sec:results}.  We discuss in Section \ref{sec:discussion} and conclude in Section \ref{sec:conclusion}.

\section{Data}
\label{sec:data}
We consider maps from the far-IR  and optical lines over a $\sim$1\arcmin~ field of view containing the central $\sim$ 5~kpc of NGC~2146 (Figure \ref{fig:fov}).  We also present new SOFIA 37~$\mu$m images to further explore the warm dust distribution.  All observations are centered on the bulge and exclude most of the extended disk.  Due to the high inclination, they do contain spiral arm regions superimposed on the bulge, as seen by the deep dust lanes in the optical image.  At the assumed distance of 17.2 Mpc \citep{Tully1988}, 1\arcsec~ corresponds to 83~pc.

\begin{figure*}[ht!]
\centering
\includegraphics[width=6.6in]{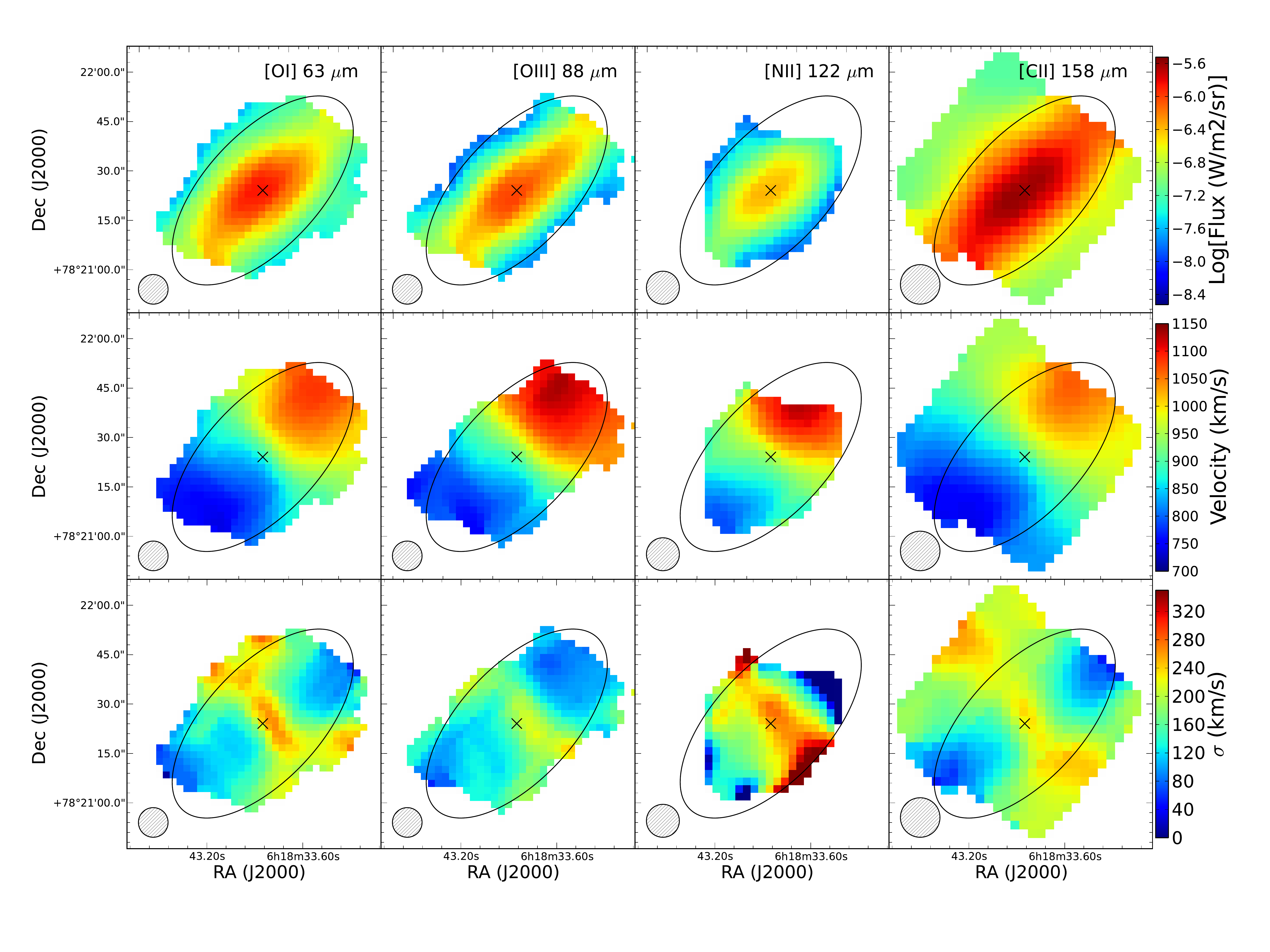}
\caption{PACS spectroscopy line maps of (left to right) [OI] 63~$\mu$m, [OIII] 88~$\mu$m, [NII]122~$\mu$m,  and [CII] 158~$\mu$m of the intensity (top), velocity (middle) and deconvolved velocity dispersion (bottom).  The ellipse and center match those shown in Figure \ref{fig:fov}. All lines are masked in regions where the signal to noise is less than five.  Uncertainties in the velocity dispersion are $\sim$5\% in the regions above and below the disk, and are much lower ($<$1\%) within the disk.  [NII] has velocity dispersion peaking at $\sim$450 \kms. The point spread function for each map is indicated in the lower left corner, and the 6 kpc ellipse shown for reference. 
\label{fig:pacs}}
\end{figure*}

\subsection{Far-IR spectroscopy}
Using the PACS instrument \citep{Poglitsch2010} onboard Herschel \citep{Pilbratt2010} we carried out far-IR spectral observations of the [OI] 63~$\mu$m,  [OIII] 88~$\mu$m, [NII] 122~$\mu$m, and [CII] 158~$\mu$m fine structure lines as part of the KINGFISH Open Time Key Program \citep{Kennicutt2011}.  The species producing [OIII], [NII], and [CII]  have ionization potentials of 35.12 eV, 14.53 eV and 11.26 eV, respectively.  Observations were centered on the galactic nucleus of NGC~2146.  All PACS spectral observations were obtained in the Un-Chopped Mapping mode and reduced using the Herschel Interactive Processing Environment (HIPE) version 11.2637.  Reductions applied the standard spectral response functions and flat field corrections, flagged instrument artifacts and bad pixels, and subtracted the dark current.  Transients caused by thermal instabilities were removed using a custom treatment designed for the KINGFISH Pipeline.  Specific information on data reduction is contained in \citet{Croxall2013} and the KINGFISH Data Products Delivery (DR3) User's Guide\footnote{http://herschel.esac.esa.int/UserProvidedDataProducts.shtml}.  Flux maps were obtained by fitting single Gaussian profiles to each projected pixel.  We extract line intensities and kinematics from a Gaussian fit to the line profile at each pixel in our line maps (Figure \ref{fig:pacs}).  Velocities are not corrected for inclination, however given the high inclination of this galaxy the corrections would be less than 10\%.  Velocity dispersion measurements have been deconvolved to correct for the instrumental broadening that ranges from 100 \kms(at [OI]) to 180 \kms(at [CII]). 
Resulting line maps achieve an angular resolution that varies by wavelength between 9\arcsec-12\arcsec.   [NII] 205~$\mu$m was also observed but is not significantly detected. 
Flux calibration of PACS data yield absolute flux uncertainties on the order of 15\% with relative flux uncertainties between each {\it Herschel} pointing within a galaxy on the order of $\sim$10\%. 

\begin{figure*}[ht!]
\centering
\includegraphics[width=6.5in]{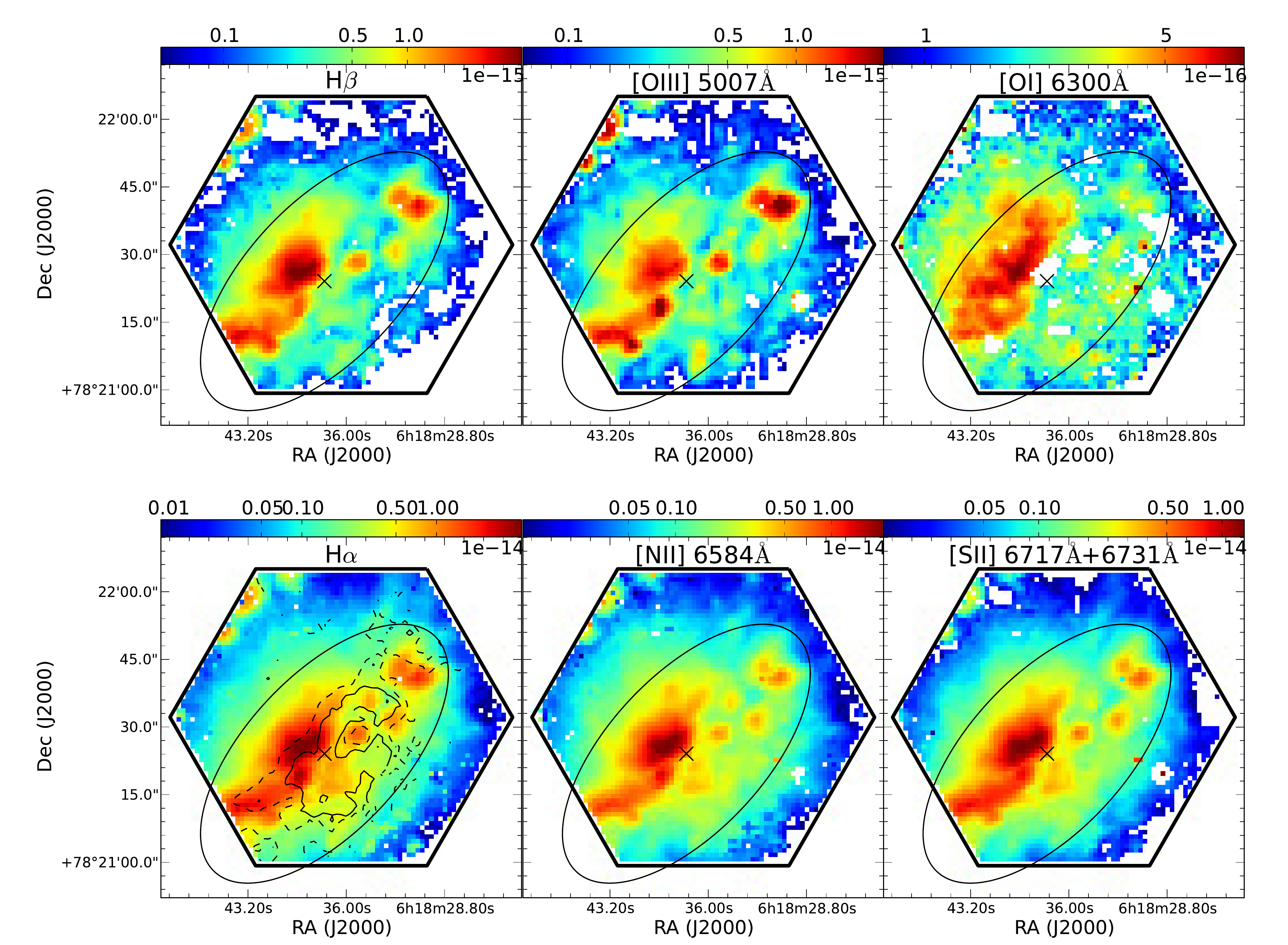}
\caption{Line maps of H$\beta$, [OIII] 5007 \AA, [OI] 6300 \AA, H$\alpha$, [NII] 6584 \AA, and [SII] 6717 \AA + 6731 \AA~ measured from the optical IFU data, where the colorbars give the line flux for each map in units of erg s$^{-1}$ cm$^{-2}$ arcsec$^{-2}$. In the H$\alpha$ image we include contours at an extinction of $A_{\rm V}$ = 3~mag (dashed) and $A_{\rm V}$ = 4~mag (solid) to indicate where the dust becomes effectively opaque in the optical.  The stellar disk and kinematic center (black) is marked for reference.
\label{fig:opticallines}}
\end{figure*}

\subsection{Far-IR imaging}
SOFIA \citep{Young2012} observations of NGC 2146 at 37.1~$\mu$m using the FORCAST instrument \citep{Herter2012} were carried out during four flights (57, 60, 63 and 64) as part of the Cycle 0 planID 81\_0059 (PI: Armus). However, due to adverse atmospheric conditions only data from three flights (57, 60 and 64) have been used in this paper. Level 3 images were retrieved from the Science Archive at the SOFIA Data Cycle System\footnote{https://dcs.sofia.usra.edu/}. Pre-processing of each image includes coadding of all chop-nod positions as well as correcting for spatial distortions. WCS is not available for Basic Science observations. Thus, the registeration of the coadded images was done by smoothing each image with a gaussian kernel and finding its centroid. Images were stacked using weights proportional to 1/$\sigma^2$, where $\sigma$ is the standard deviation of the sky around the galaxy. Astrometry was applied by comparing the centroid to the MIPS \citep{Rieke2004} 24~$\mu$m image.  Flux-calibration was performed using the multiplicative factor CALFACTR available in the header of each image.  Final image resolution is diffraction-limited with a full width at half maximum of 3\farcs5.

We also present archival Spitzer MIPS 24 $\mu$m \citep{Armus2009} and Herschel PACS 100 $\mu$m, 160 $\mu$m and
SPIRE 250 $\mu$m \citep{Kennicutt2011} images for comparison.

\subsection{Optical IFU data}
Optical integral field unit (IFU) data were taken at the Calar Alto 3.5m telescope using the PMAS instrument in PPAK mode \citep{Roth2005, Kelz2006}, which consists of 331 fibers each 2\farcs68 in diameter arranged in a hexagonal pattern to cover a $\sim$1\arcmin~field of view.  Observations were performed using the V300 grating, which covers 3700~\AA~to 7000~\AA~at $\sim$180~\kms~velocity resolution, and dithered in three positions to recover the full flux and subsample the fiber size.  The data reduction has been described in detail  in  \cite{Kreckel2013}.  All spectra were reduced using the p3d package \citep{Sandin2010}, and line fluxes were extracted using the Gandalf \citep{Sarzi2006} and pPXF \citep{Cappellari2004} software packages.   We present line maps with a 2.5\arcsec~ spatial resolution for the H$\beta$, [OIII] 5007~\AA, [OI] 6300~\AA,  [NII]  6584~\AA, H$\alpha$ and both [SII] 6717~\AA~and 6731~\AA~lines in Figure \ref{fig:opticallines}. The [NII] 6548~\AA~line is also detected and is fit assuming a fixed ratio with the [NII] 6548~\AA~line.  The emission of [OI] is redshifted away from the skyline contamination.

\begin{figure*}[ht!]
\centering
\includegraphics[width=7.1in]{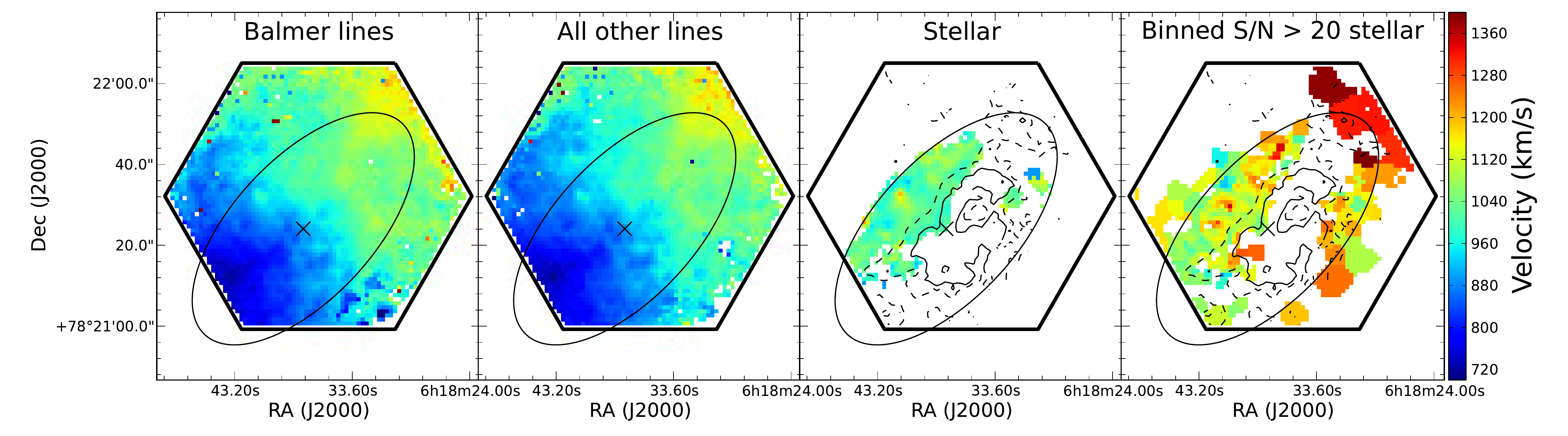}
\caption{From left to right:  Velocities  from fitting together all Balmer lines (H$\alpha$, H$\beta$, and where detected H$\gamma$, H$\delta$), all other PPAK emission lines ([OIII], [OI], [NII], [SII]),  from the stellar absorption kinematics at full resolution and from the stellar absorption kinematics after binning pixels to a signal to noise $>$ 20.  Emission line maps are masked at an amplitude over noise of 2, stellar kinematics are masked at a signal to noise of 5.  Contours show a measured extinction of $A_{\rm V}$ = 3~mag (dashed) and $A_{\rm V}$ = 4~mag (solid) to indicate where the dust becomes effectively opaque in the stellar continuum.  Velocity dispersions are not resolved due to the instrumental dispersion. The stellar disk and kinematic center (black) is marked for reference.
\label{fig:optkinematics}}
\end{figure*}

When fitting the lines and continuum we tie the velocities and dispersion of the Balmer lines (H$\alpha$, H$\beta$, and where detected H$\gamma$, H$\delta$) and separately tie together the kinematics of the remaining nebular lines ([OIII], [OI], [NII], [SII]). Our requirement that tied lines exhibit matched dynamics is based on the assumption that they are coming from the same regions within the galaxy.   Tying the lines improves the fit for low signal to noise lines, allowing us to measure the velocity centroid more accurately (Figure \ref{fig:optkinematics}), despite the limited instrumental spectral resolution. We then obtain two independent measures of the kinematics from the Balmer and forbidden lines.  As with the far-IR lines, velocities are not corrected for inclination.  We measure the line centroid to within 20~\kms~but do not resolve the velocity dispersion due to the large $\sim$180~\kms~instrumental spectral resolution.  

Simultaneous with our emission line fit using Gandalf we also fit with pPXF a combination of stellar templates based on the \cite{Tremonti2004} adaptation of the \cite{Bruzual2003} simple stellar population (SSP) models.  These cover a range of ages (5 Myr- 11 Gyr) at solar metallicity.  The wide wavelength range of our spectra and the fact that it includes the age-sensitive 4000\AA~ break allows us to break the dust-age degeneracy (see also \citealt{Kreckel2013}).    From this stellar template fit, which is masked to exclude emission lines and includes numerous stellar absorption features, we recover a stellar kinematic fit including both a velocity centroid and dispersion.  Stellar velocities are accurate to $\sim$50~\kms~and stellar velocity dispersions are largely unresolved at our spectral resolution.  
These measurements are only possible in regions not suffering from high dust extinction, therefore for a wide region located north-east of the dust lane we obtain robust solutions.

Voronoi binning \citep{Cappellari2003} of neighboring spectra to ensure a minimum signal to noise of 20 per region in the stellar continuum recovers detections across much of the stellar disk (Figure \ref{fig:optkinematics}). 
For these binned regions we fit a non-linear combination of 330 stellar templates from the Indo-U.S. library \citep{Valdes2004}. This subset was chosen to cover in a uniform manner stellar parameters like effective temperature, surface gravity, and metallicity. The choice of stellar templates, rather than SSP templates, results in a slightly better fit to the H$\beta$ absorption feature and serves as a check that our choice of stellar templates is not biasing our results. To derive the mean stellar velocity and velocity dispersion we again use the pPXF method \citep{Cappellari2004}, and derive errors via Monte Carlo simulations. More details about these steps to extract stellar kinematics are given in Falcon-Barroso et al. (in prep).  We observe very different stellar kinematics compared to the ionized gas, which we discuss in more detail in Section \ref{sec:kin}.

\section{Results}
\label{sec:results}

While H$\alpha$ is a well established tracer of the ionized gas in galaxies, with the far-IR fine structure lines we have the opportunity to trace the ionized gas through the [OIII] 88~$\mu$m and  [NII] 122~$\mu$m emission without the high dust extinction suffered in the optical.  Further, the [OI] 63~$\mu$m line allows us to trace the neutral atomic gas, and [CII] 158~$\mu$m traces both the neutral atomic gas and the ionized medium.  

In Figure \ref{fig:pacs} we see that all far-IR line emission is strongly peaked towards the galaxy center, with the extent very well matched to the optical disk.  The center of the line emission agrees within the astrometric uncertanties to the center seen in the stellar mass as traced by the 3.6~$\mu$m emission (Figure \ref{fig:fov}) and with the kinematic center determined from the CO velocity field \citep{Tsai2009}.  We see clearly the benefit of long wavelength observations, as the center imaged in the optical is obscured by a strong dust lane (Figures \ref{fig:fov} and \ref{fig:opticallines}).  

In galactic winds, neutral and ionized gas emission originates mainly from within the walls of the outflow cone \citep{Veilleux2005}. This material is expected to be turbulent and shocked due to interactions with the hot wind, resulting in increased excitation in the gas.  Emission from opposite sides of the cone can produce line splitting, when seeing the opposing walls together in projection, however, in NGC~2146 we expect the geometry will make this difficult to observe.  
The parts of the outflow cone we see are essentially perpendicular to the line of sight, with the opposing walls of the cone well separated in projection.  
Broadened emission lines provide an additional outflow signature, and can also arise from turbulence in the emitting material within the cone walls.  In NGC\,2146, outflow signatures are observed both in the line kinematics and the gas excitation as we discuss below.

\subsection{Kinematic signatures}
\label{sec:kin}
\subsubsection{Velocity dispersion}
\label{sec:veldisp}
In NGC~2146, an increased velocity dispersion is observed in all of the far-IR emission lines (Figure \ref{fig:pacs}) both along the minor axis and in regions offset above and below the disk. The different lines show some variation in the line dispersion, broadly consistent with what is expected as they originate from different gas phases.  The [NII] emission displays the largest velocity dispersion, consistent with shock excitation affecting a large volume of low ionized gas.  The [CII] emission arises from both the ionized and neutral phases, and the overall agreement with the neutral gas traced by [OI] suggests that the neutral phase may be dominating the emission mechanism.  As [OIII] arises from higher ionization material, this could account for the lower dispersion observed.  All four lines show very similar morphology, providing a consistent picture of the outflow. 

For the remainder of our discussion, we focus on only [CII] and [OI], as [CII] has the highest signal to noise but [OI] has better instrumental velocity resolution. We observe a deconvolved dispersion of 250~\kms~in the [CII] and [OI] lines, which reflects an increase of 50-100~\kms~above the velocity dispersion measured in the disk.  Multiple components may be present but are not resolved at the $\sim$100~\kms~ instrumental resolution, and all line emission along the minor axis is centered at the systemic velocity (Figure \ref{fig:pacs}).  This is fully consistent with the previously established picture of the outflow geometry that results in an alignment of the far side of the cone to the north and the near side of the cone to the south perpendicular to the line of sight \citep{Greve2000, Tsai2009}, with the line emission from the opposing cone walls overwhelmed in projection by the bright central disk.

\begin{figure*}[ht!]
\centering
\includegraphics[width=7in]{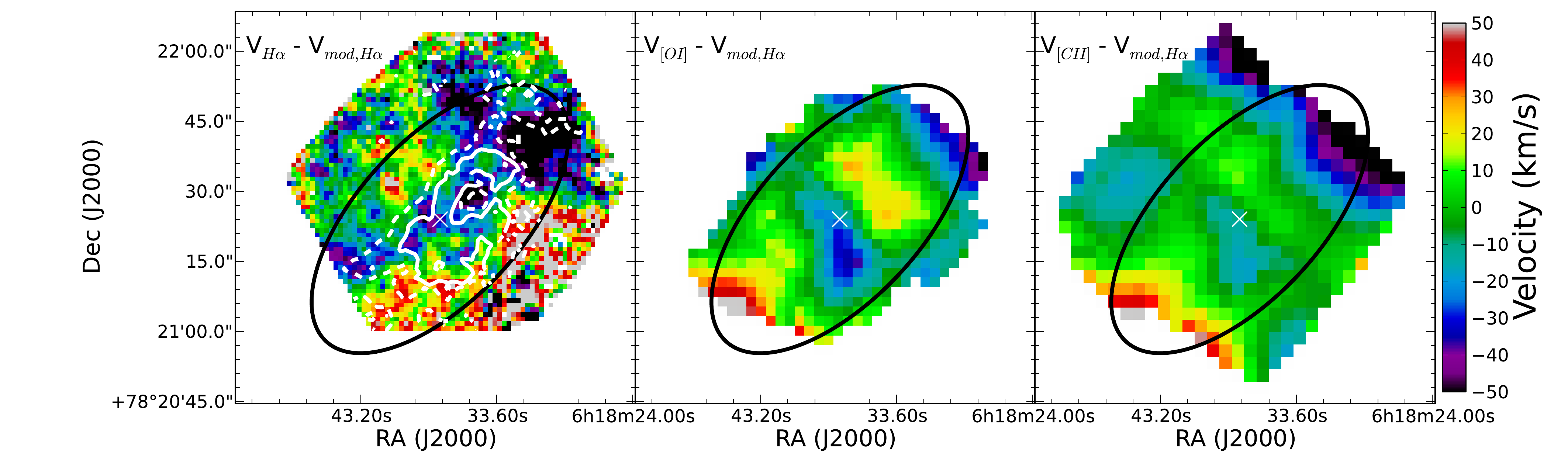}
\caption{Residuals velocity fields in  H$\alpha$ (left),  [OI] (center) and  [CII] (right) subtracting a model velocity field based on the H$\alpha$ rotation curve \citep{Greve2006}.  Regions in the H$\alpha$ map along the disk major axis that show blueshifted components compared to disk rotation anti-correlate with regions of high extinction as judged from the Balmer decrement (outlined in white at $A_{\rm V}$ = 3~mag (dashed) and $A_{\rm V}$ = 4~mag (solid)), suggesting these correspond to H~\textsc{ii} regions foreground to the dust lane that runs along the major axis (also visible in Figure \ref{fig:opticallines}).  Along the major axis of the PACS maps we see irregularities in the kinematics at the disk center.  Above the disk there is a patch of blueshifted emission and below an extended region of redshifted emission as seen in the H$\alpha$ velocity field residuals.  There is no corresponding shifted emission in the PACS line maps.
\label{fig:resid}}
\end{figure*}

This increase is well in excess of the increased velocity dispersion that results from beam smearing due to low spatial resolution observations of an edge-on disk (e. g.  \citealt{Tamburro2009, Epinat2010}).  We model this effect in our far-IR observations by constructing a two dimensional exponential disk model with 63$^\circ$ inclination, to which we assign a velocity at each position based on the H $\alpha$ rotation curve model of \cite{Greve2006}.  From this we construct a data cube with flux in a single channel (corresponding to an intrinsic $\sigma$ = 0 \kms), smooth this to the $\sim$0.8 kpc resolution of our observations, then fit a Gaussian profile to each position in the disk.  The resulting increase in central velocity dispersion reaches $\sim$80 \kms over a narrow 0.5 kpc region along the major axis and a more extended 2.5 kpc region along the minor axis.  This is  substantial, but significantly less than the $\sim$250~\kms~velocity dispersion that we measure.

\cite{Tsai2009} also found evidence in the CO kinematics for a nuclear disk with a higher rotation velocity extending 300 pc (4\arcsec) in radius. Such a nuclear disk would be unresolved in our observations, but could potentially contribute to beam smearing in the center.  However, given the uniformity in the increased dispersion both along the minor axis as well as out in the wind region, we believe that this indicates that we truly are tracing the wind back to the central wind-launching region.  This is very similar to what is seen in M82 \citep{Contursi2013}.  In NGC 2146 the region of increased dispersion is not resolved in [OI], but the area with dispersions above 200 \kms extends less than 1 kpc at the disk center and is slightly offset to the north of the disk kinematic center but agrees within the PACS astrometric uncertainty.  The outflow is cylindrical  out to approximately 500 pc, and broadens beyond this point into conical outflows. The outflow extends to the edge of the field of view, approximately 2.5 kpc above the disk, and presumably even further above the disk. The region north-east of the disk shows an opening angle of $\sim$90$^\circ$, to the south-west the opening angle is much broader, approximately 120$^\circ$.

Although our optical spectral resolution is comparable to that of [CII], we do not observe large 250~\kms~velocity dispersions in H$\alpha$. However due to the overall lower signal to noise of the optical emission lines, we also can not reliably decompose velocity dispersions in the optical below 250~\kms~ from the instrumental velocity dispersion.  
We do see increased velocity dispersion in the [NII] lines (at least a factor of two or more above the disk value up to about 250 \kms) to the south-west in the area of the outflow.  
The outflow can also be considered in light of higher velocity resolution Fabry-Perot measurements of the H$\alpha$ line emission in this galaxy as part of the GHASP survey \citep{Epinat2008}.  These data clearly show multiple components, a detailed breakdown of which is beyond the scope of this paper but has been investigated in detail using long slit spectra \citep{Armus1995, Greve2000}.  This supports the idea that a complex structure exists in the emitting material.

\subsubsection{Residual velocity field}
\label{sec:residvelfield}
To search for further kinematic evidence of outflows we have constructed a 2D model velocity field from the rotation curve measured by \cite{Greve2006} from longslit H$\alpha$ kinematics. We considered also the rotation curve measured by \cite{Tsai2009} based on the CO velocity field, however it does not extend far enough radially to include the flattened part of the rotation curve and provides a nearly identical picture of the non-circular motions.  The H$\alpha$ rotation curve extends to larger radii, although extinction may bias the determination of the line-of-sight velocity centroid.  We subtract this model velocity field from our Balmer line,  [OI] and [CII] velocity fields and consider the residual velocity fields (Figure \ref{fig:resid}).  

\begin{figure*}
\centering
\includegraphics[width=3in]{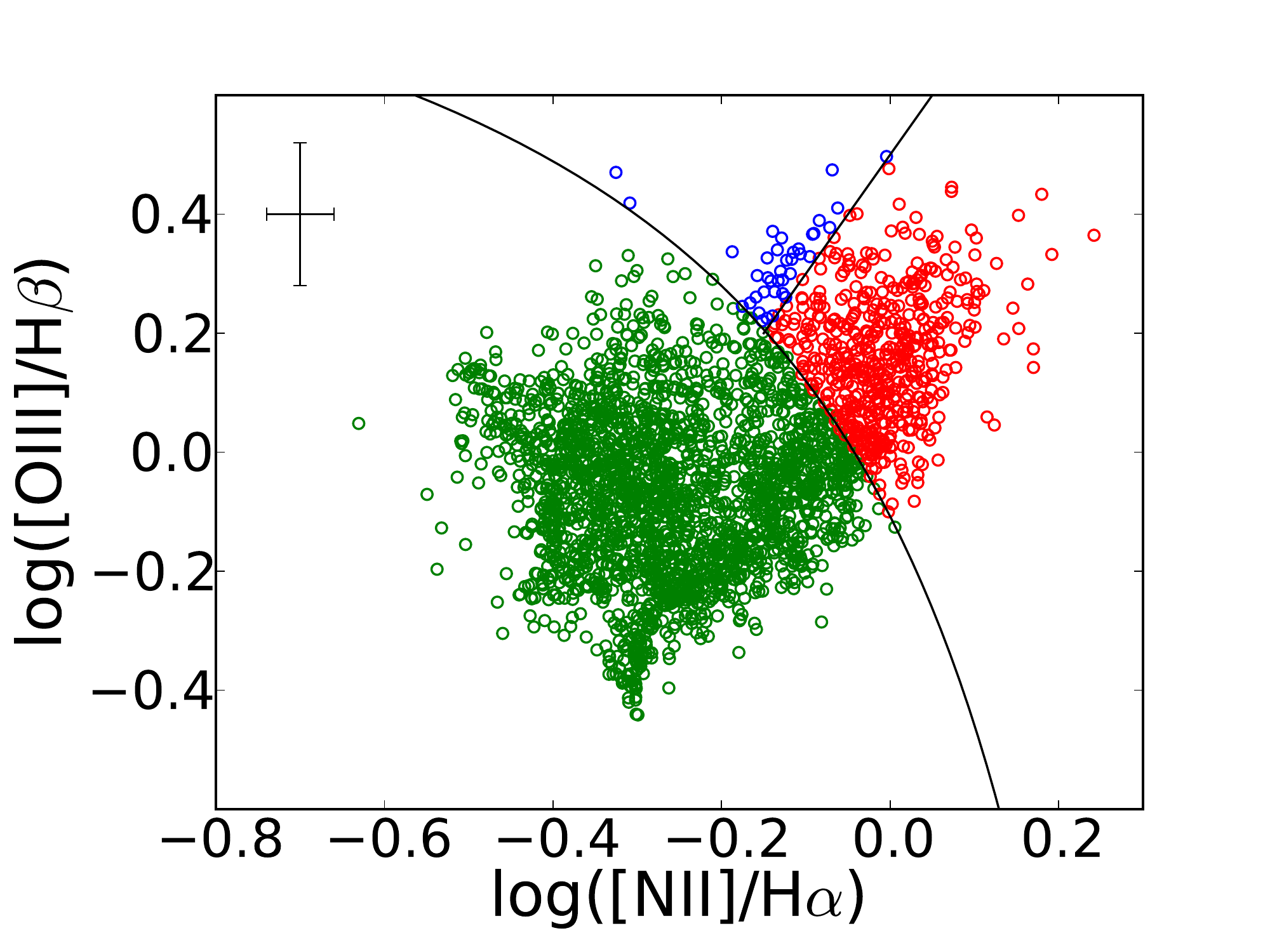}
\includegraphics[width=3in]{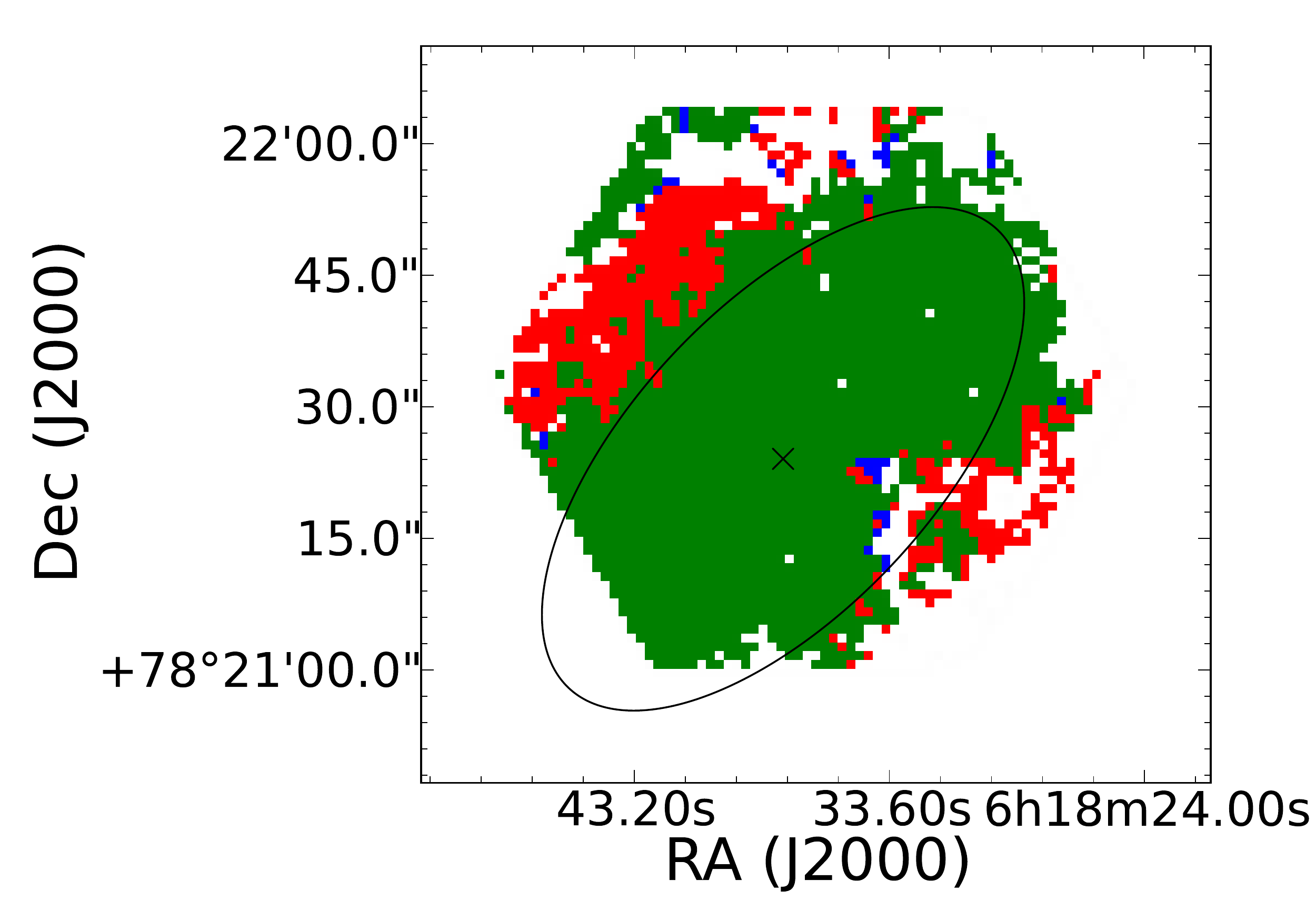} \\
\includegraphics[width=3in]{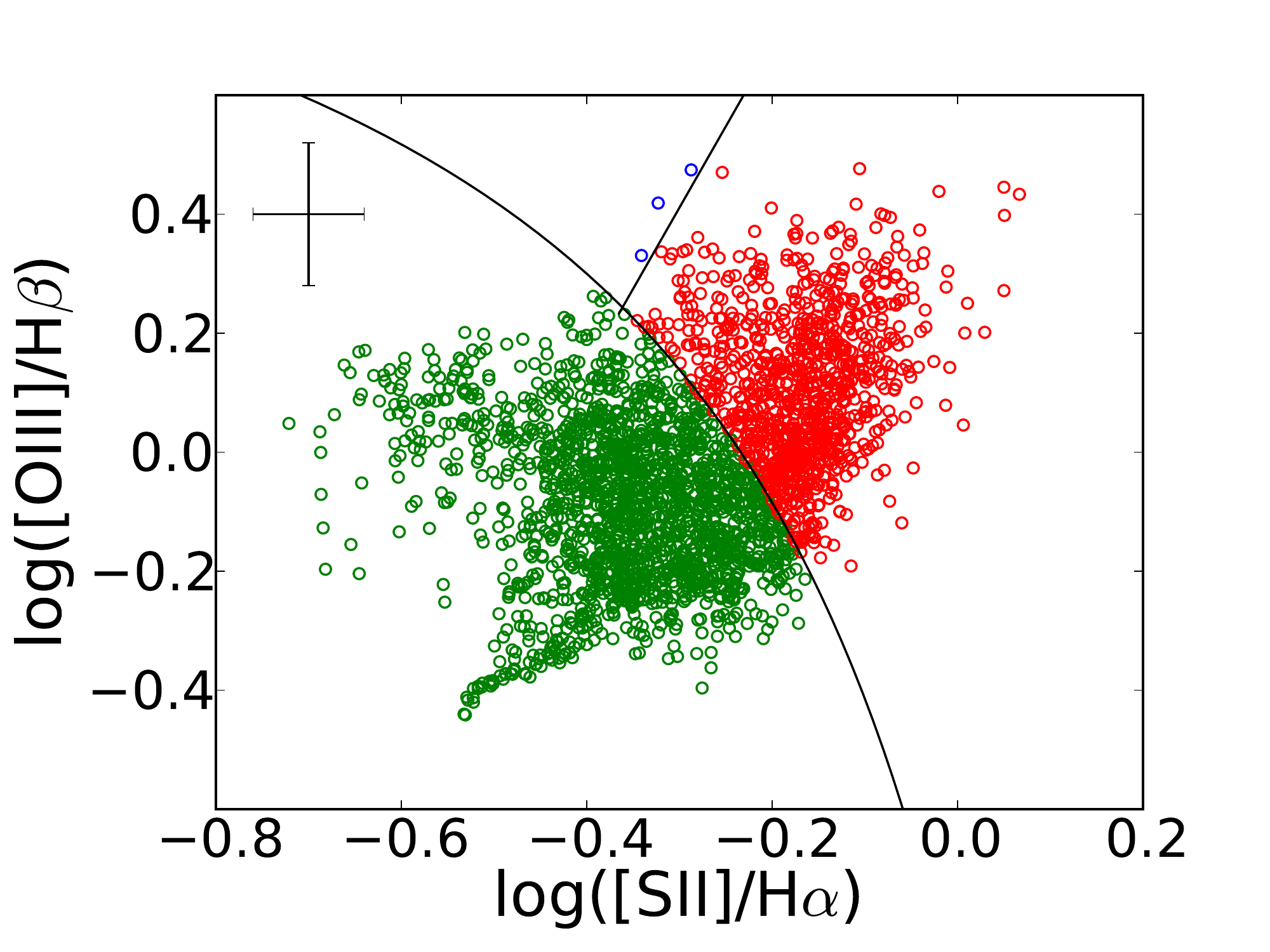}
\includegraphics[width=3in]{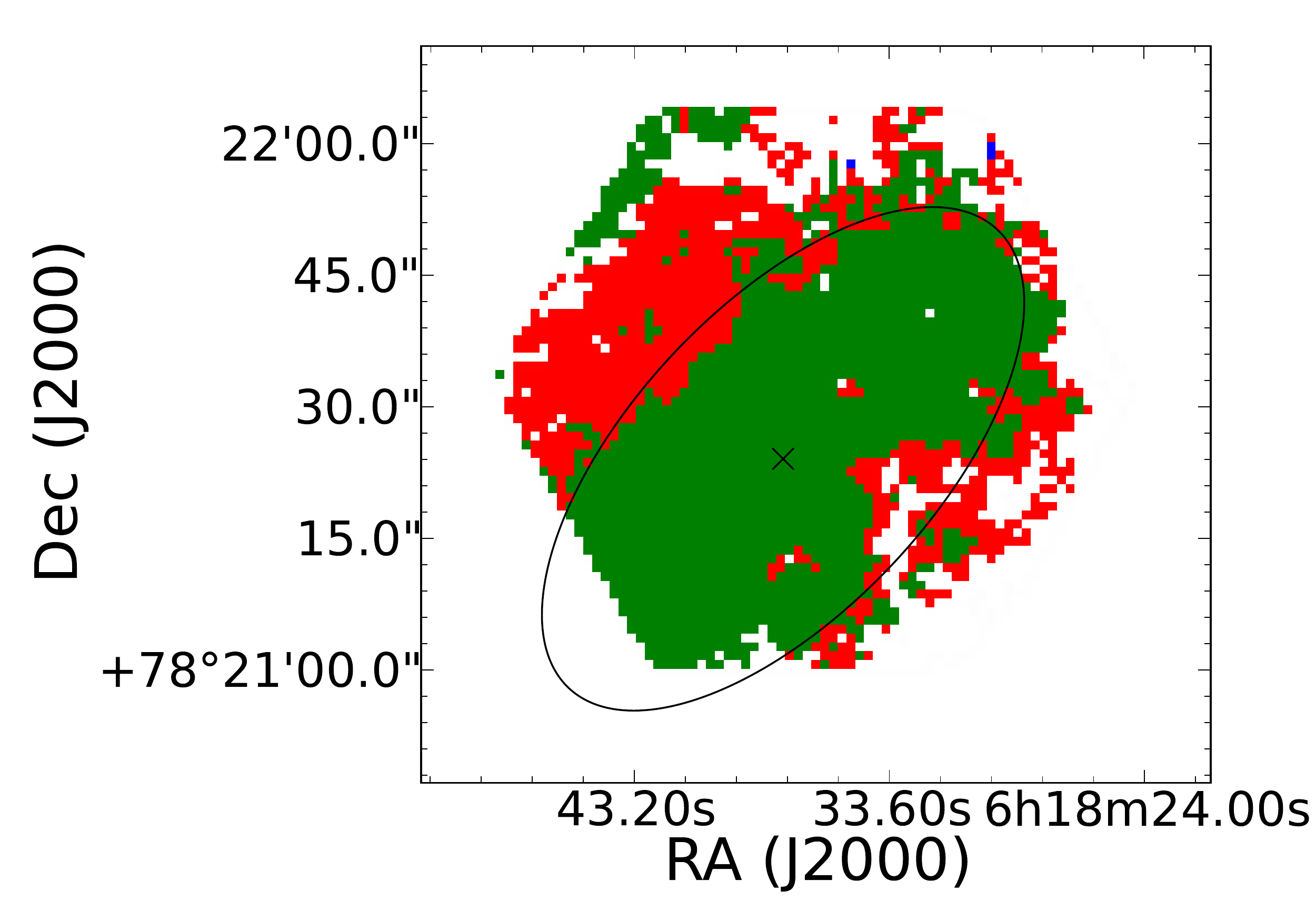} \\
\includegraphics[width=3in]{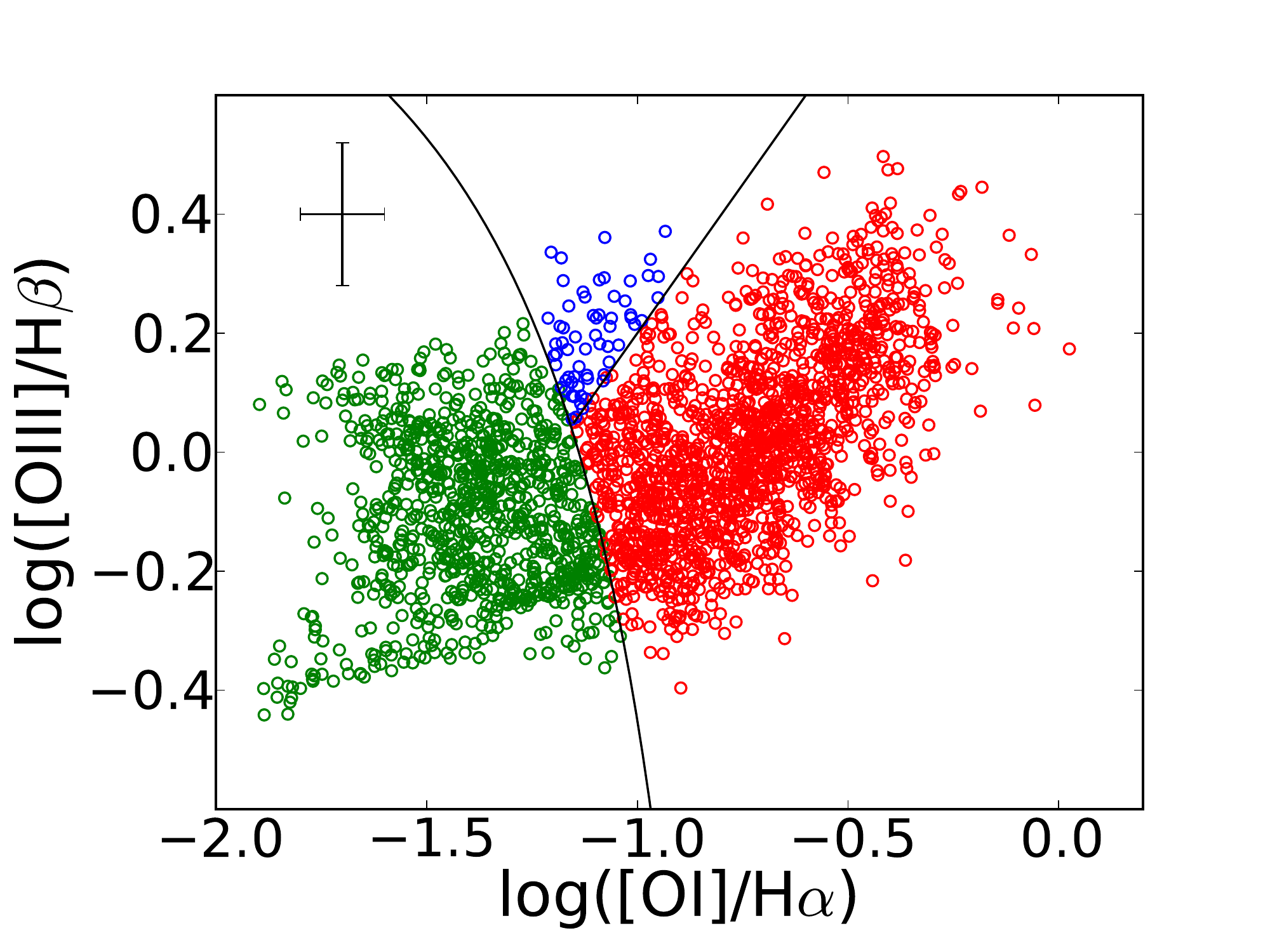}
\includegraphics[width=3in]{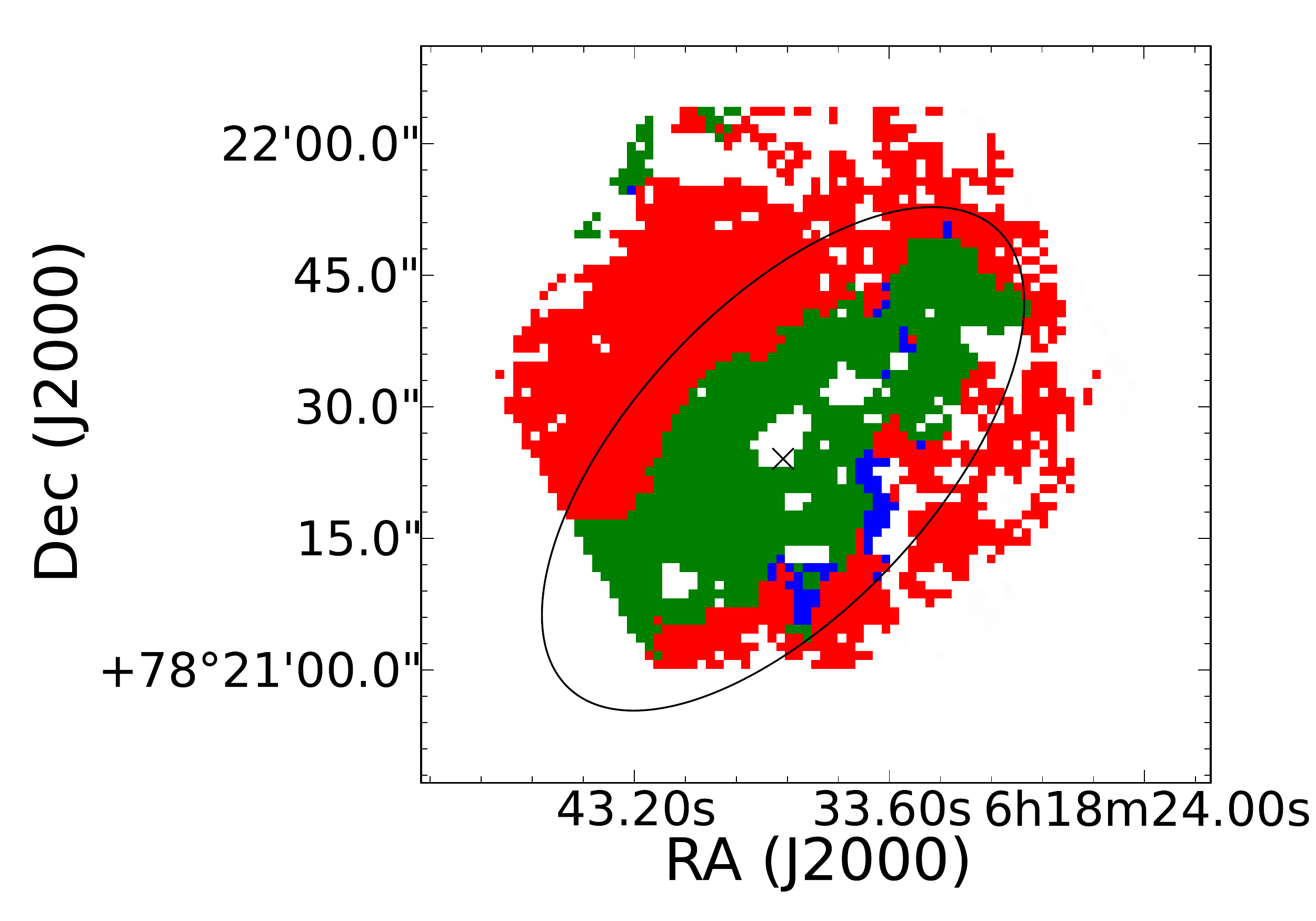}
\caption{Left: Spatially resolved BPT diagrams for three optical diagnostic tracers ([NII]/H$\alpha$, [SII]/H$\alpha$, [OI]/H$\alpha$).  
We see enhanced [OI], [SII] and [NII]/H$\alpha$ line flux ratios, beyond what is expected given an extreme starburst limit (curved line, \citealt{Kewley2001}) and excluded by fiducial AGN indicators (straight line, \citealt{Sharp2010}).  Right: Corresponding maps of the color-coded regions, as identified in the figures at left.   The stellar disk and kinematic center (black) is marked for reference.
\label{fig:bpt}}
\end{figure*}

The residual velocity fields show extended blueshifted regions along the major axis that correspond to the position of bright HII regions foreground to the dust lane (Figure \ref{fig:resid}, left), as seen in the H$\alpha$ line maps.  Similar to results presented in \cite{Armus1995}, the residual velocity field also shows blueshifted emission to the north-east of the disk and patchy redshifted emission to the south-west that are separated by about 100~\kms.  This emission was not detected by \cite{Greve2000}, however it is clear in our maps that this detection would be extremely dependent on the exact slit position.   The redshifted region corresponds approximately to the southern shock-excited region (see Section \ref{sec:excitation} and Figure \ref{fig:bpt}). The extent of this region agrees well with the position of the anomalous CO clumps identified by \cite{Tsai2009}, however they find much lower redshift velocities of $\sim$10~\kms.  The blueshifted region is much more compact than the shocked region, and aligns with the western edge of the molecular superbubble identified by \cite{Tsai2009} (see also Section \ref{sec:coout}).  Both the redshifted and blueshifted regions appear to originate offset from the disk minor axis.

The PACS line residuals are large along the disk major axis (Figure \ref{fig:resid}, right).  This may be evidence that our model is missing a nuclear disk component, previously observed in the CO kinematics \citep{Tsai2009}, however with a 300~pc radius that disk should be unresolved in both [OI] and [CII].  Another cause could be that we have neglected any radial motions in the disk, which may be expected  since this starburst is presumably fed through significant gas inflow.  However, the residuals from modeling the CO disk by \cite{Tsai2009} show no such patterns.  Neither [CII] nor [OI] show evidence for significant redshifted or blueshifted material, particularly in the regions where this is seen in the H$\alpha$ kinematics. It is possible the H$\alpha$ emission comes from clumpy features associated with the foreground and background spiral arms, while the far-IR lines, unbiased by extinction, mainly trace gas in the main disk and outflow.  

\subsection{Diagnostic line ratios}
\label{sec:excitation}
Optical line ratios can be used to establish the ionization and excitation state of gas in galaxies \citep{Baldwin1981, Kauffmann2003}, though they have most commonly been applied to either integrated galaxy spectra or nuclear spectra when trying to distinguish nuclear excitation sources (SF, LINER, AGN).  However,  similar physics applies in resolved regions within a galaxy, where the excitation sensitive indicators [NII]/H$\alpha$, [SII]/H$\alpha$, and [OI]/H$\alpha$ should be enhanced in regions of the outflow, as expected from shocked gas. Due to the high signal to noise the [NII]/H$\alpha$ map is the most complete, although due to the high inclination and significant attenuation we still may not probe that far into the disk along the major axis.  The density diagnostic [SII]$\lambda$6717/[SII]$\lambda$6731 assuming a gas temperature of 10,000 K shows a fairly uniform density of n$_e$ $\sim$100 cm$^{-3}$.

Evidence for shock excitation along the minor axis of NGC 2146 has been previously identified through optical excitation diagnostics \citep{Armus1995, Hutchings1990, Greve2000}.  Here, combining line ratio maps pixel by pixel to construct BPT diagnostic diagrams \citep{Baldwin1981}, we see increased flux in the low excitation lines ([NII], [OI], [SII]) compared to H$\alpha$ throughout this central region (Figure \ref{fig:bpt}, left).  These ratios are selected to involve lines nearby in wavelength, such that reddening due to dust extinction will not be significant and can be neglected.  In all three tracers, we see evidence for high excitation (in red)  beyond what is expected given an extreme starburst limit (curved line, \citealt{Kewley2001}) and excluded by fiducial AGN indicators (straight line, \citealt{Sharp2010}).  By restricting ourselves to only high confidence emission line detections we have ruled out any bias due to stellar absorption corrections or other systematic errors in these lower surface brightness regions.  

Mapping these regions to their corresponding physical locations within the galaxy (Figure \ref{fig:bpt}, right), the high excitation regions appear located $\sim$2 kpc above and below the galaxy disk, and extend $\sim$5 kpc horizontally above the disk all the way to the edge of the field of view, and is consistent with shock excitation.  A combination of star forming and shocked regions along the line of sight would likely mask any shocked emission closer to the disk.  Shock excitation of this sort above the disk has often been seen in longslit studies of starburst-driven galactic superwinds \citep{Armus1989, Heckman1990}, and our map appears very similar to the starburst  driven winds studied by \cite{Sharp2010} in their study of 10 galaxies with galactic winds using IFS observations.   No high ionization nuclear source is seen in the  [OIII]/H$\beta$ data, which is unsurprising as the nuclear region is highly attenuated.

\begin{figure}[b!]
\centering
\includegraphics[width=3.3in]{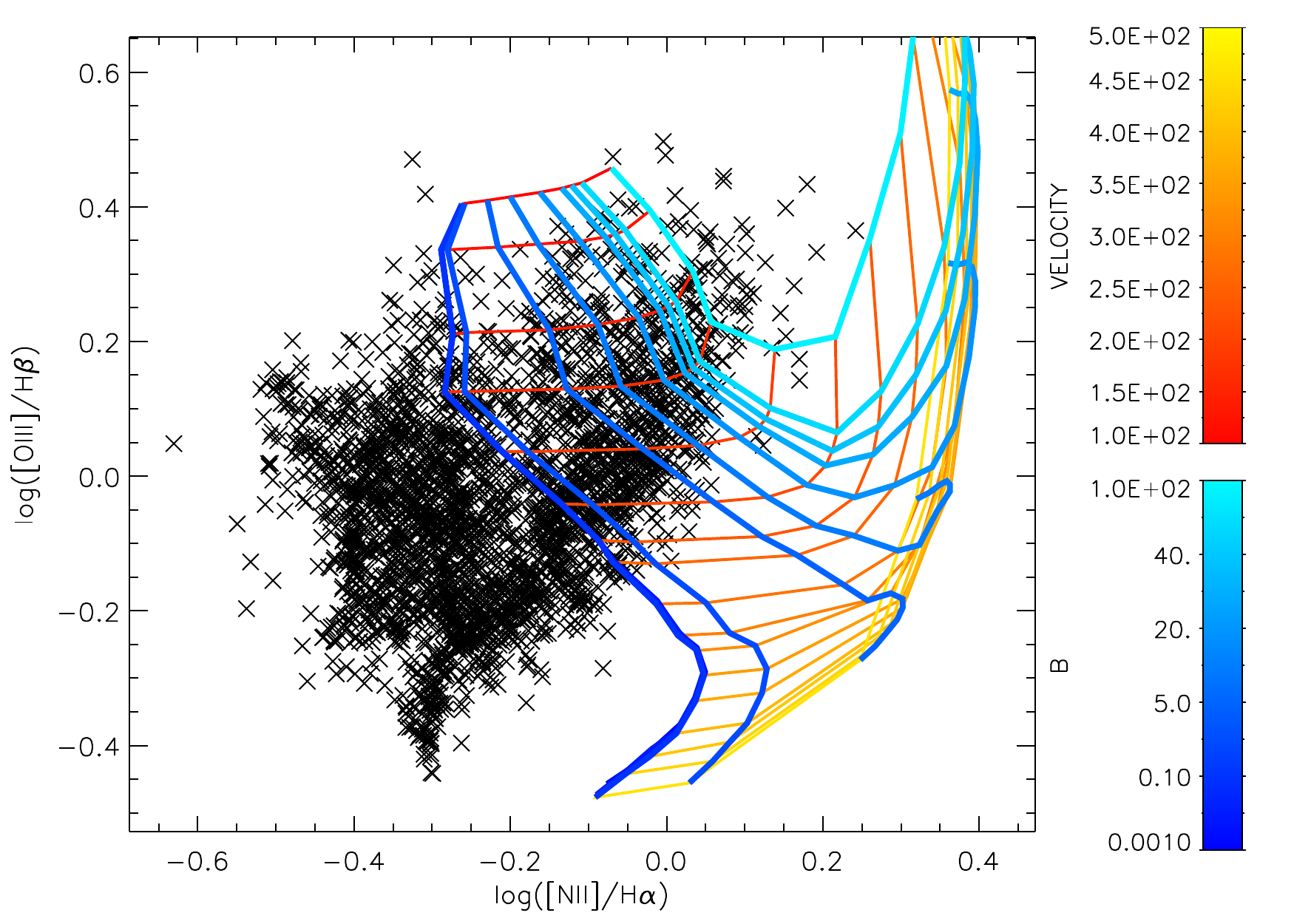}
\caption{Shock modeling with solar metallicity models assuming T$_e$ = 10,000 K and n$_e$=100 cm$^{-3}$ (top) for the resolved BPT diagrams shown in Figure \ref{fig:bpt}. Model grid parameters are the shock velocity and B, the transverse component of the preshock magnetic field \citep{Allen2008}. Modeled shock velocities within NGC~2146 range from 100-200 \kms, though we note that the contribution from photoionized regions that fall along the line of sight will bias us towards lower shock velocities.  
\label{fig:shocks}}
\end{figure}

\begin{figure*}[ht]
\centering
\includegraphics[width=6.3in]{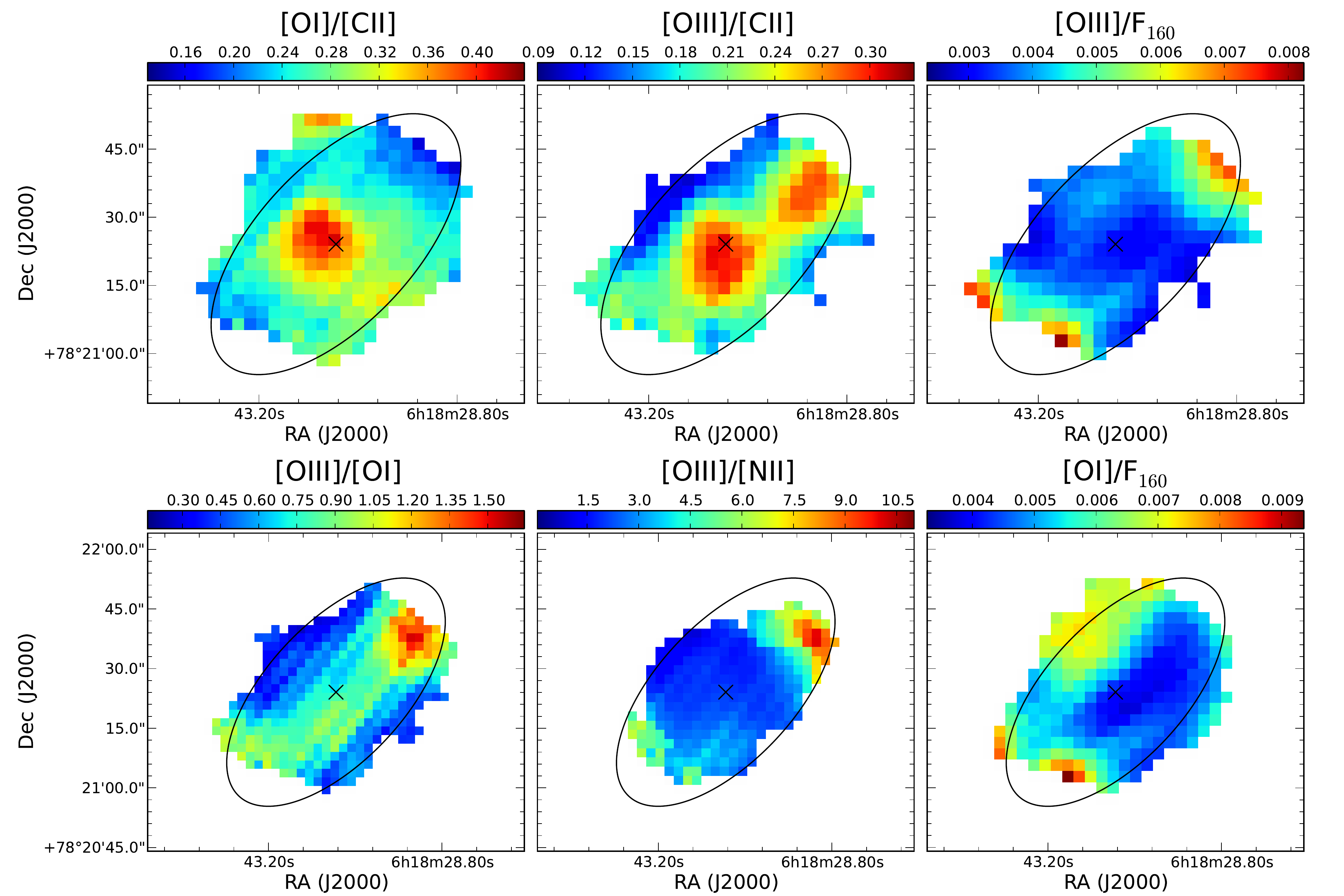}
\caption{Line ratios of the far-IR fine structure lines.  The [OIII]/[CII] and [OIII]/[OI] ratios are consistent with star formation within the disk.  The [OI]/[CII] ratio shows a different morphology, with a bright nucleus and a structure of high [OI]/[CII] that lies perpendicular to the disk, tracing the outflow.  Uncertainties are typically 5\%.  The stellar disk and kinematic center (black) are marked for reference.
\label{fig:PACSlineratios}}
\end{figure*}

\begin{figure*}[h!]
\centering
\includegraphics[width=6.5in]{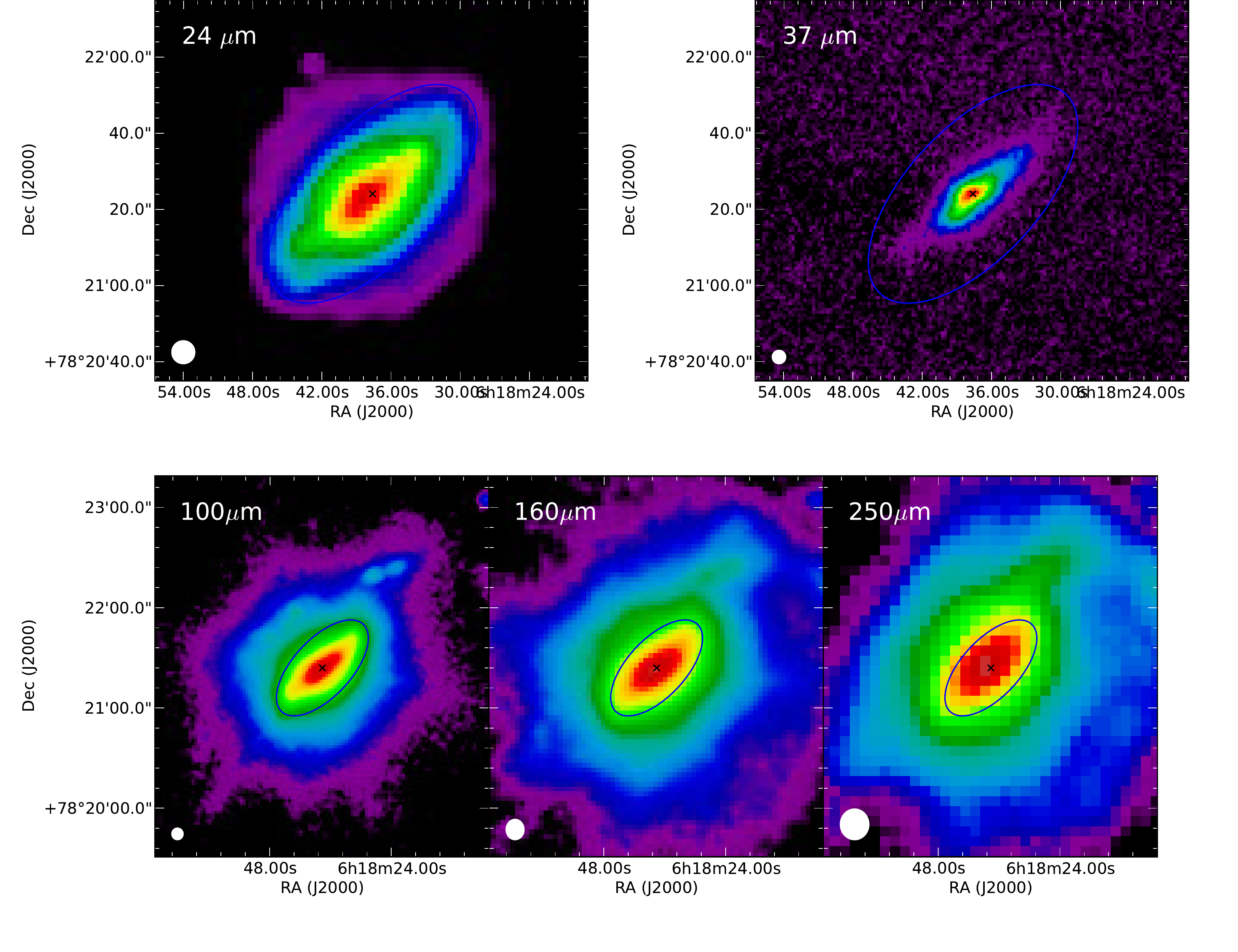}
\caption{Dust emission in NGC~2146.  Warm dust is traced by the Spitzer MIPS 24~$\mu$m emission (top left) and SOFIA FORCAST 37~$\mu$m emisison (top right).  Cold dust is traced by the Herschel PACS 100~$\mu$m (bottom left), 160~$\mu$m (bottom center), and SPIRE 250~$\mu$m (bottom right) emission.  The emission generally follows the disk of the galaxy, with extended features that trace the optical spiral arms seen extending beyond the field of view of our line maps.  Hexagonal features in the PACS  images at low flux levels are due to the instrumental point spread function.  The stellar disk (blue oval) and kinematic center (cross) is marked for reference and is the same in all figures.
\label{fig:spire250}}
\end{figure*}

\begin{figure*}
\centering
\includegraphics[width=2.2in]{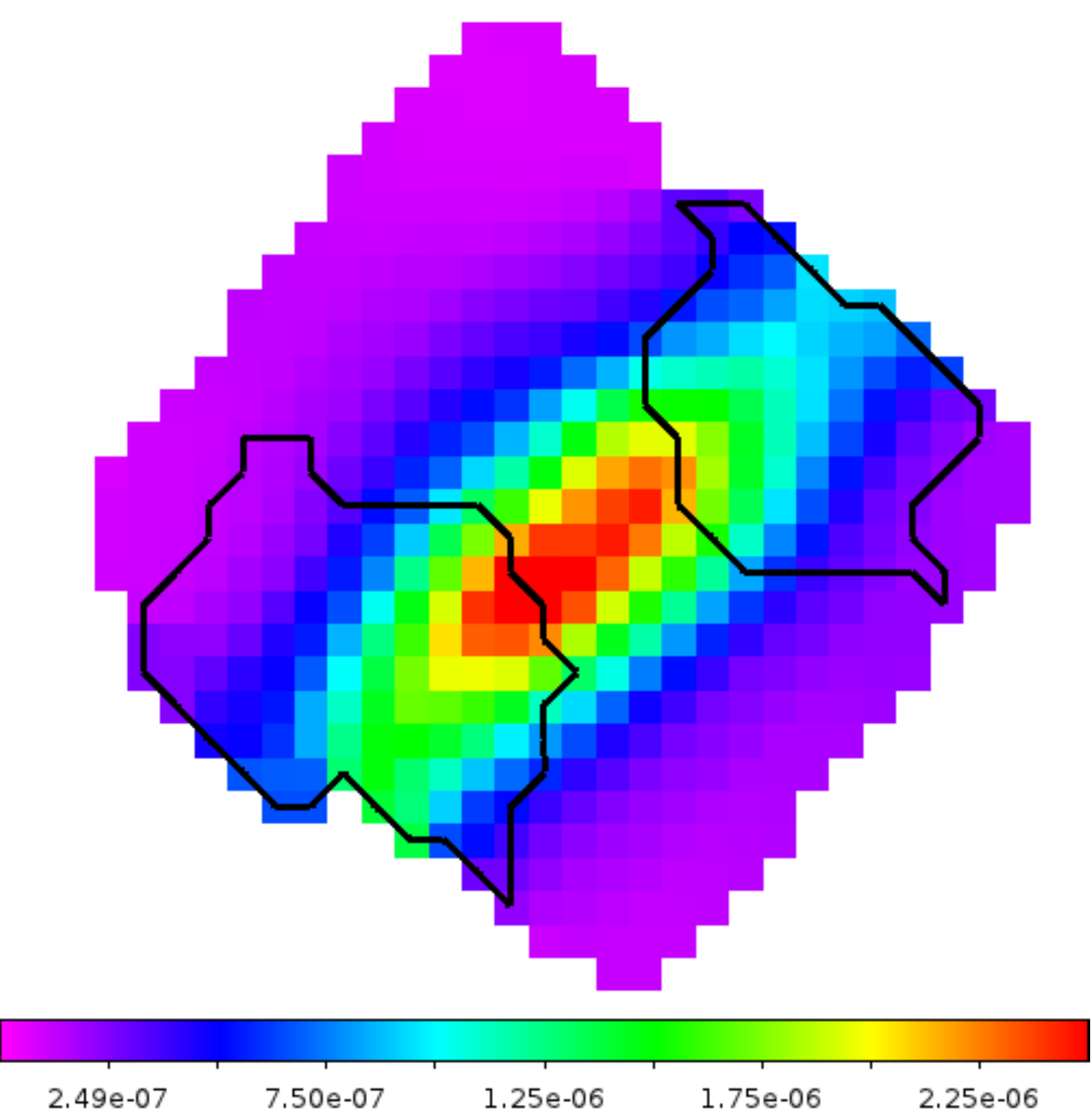}
\includegraphics[width=2.2in]{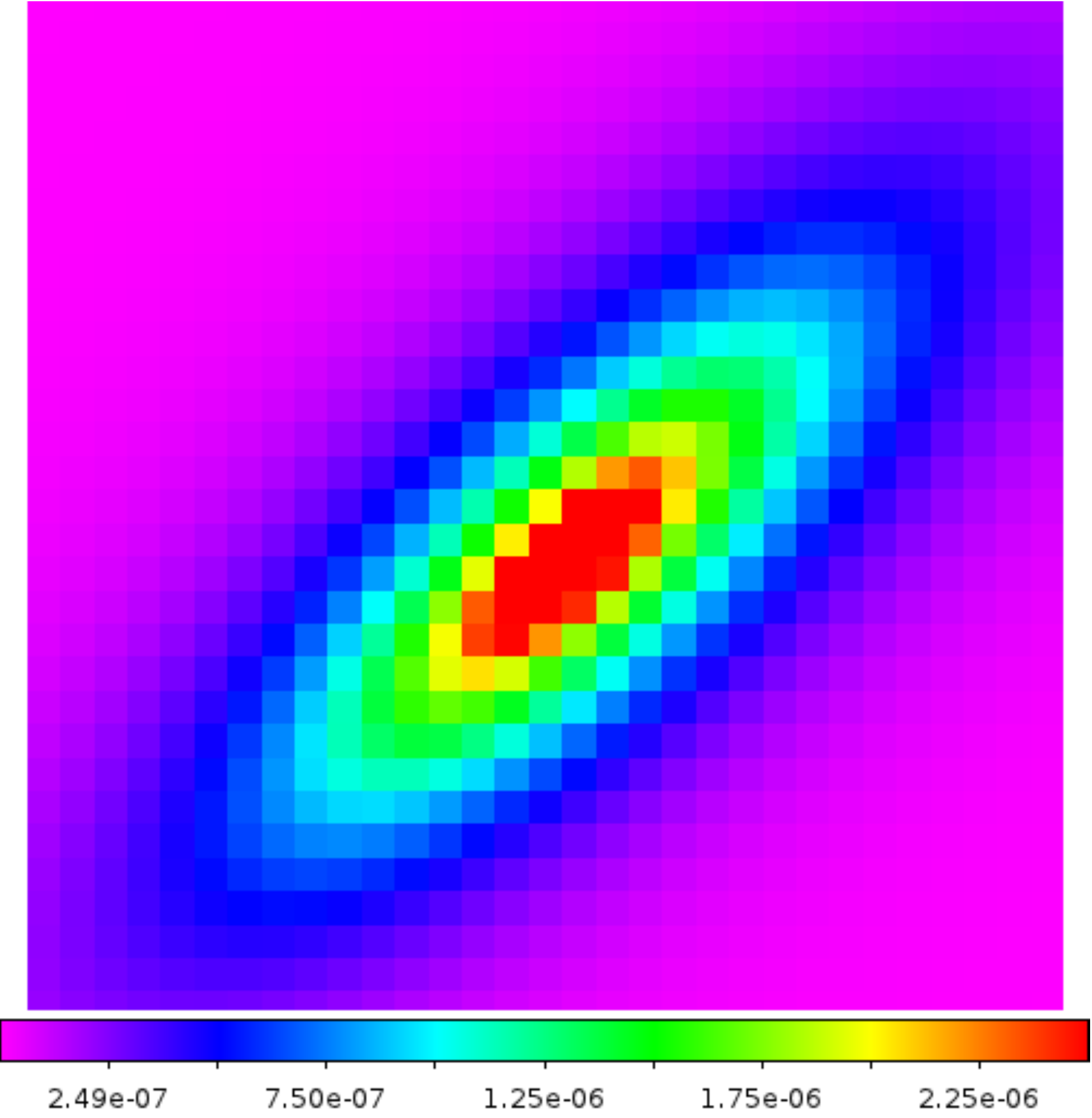}
\includegraphics[width=2.2in]{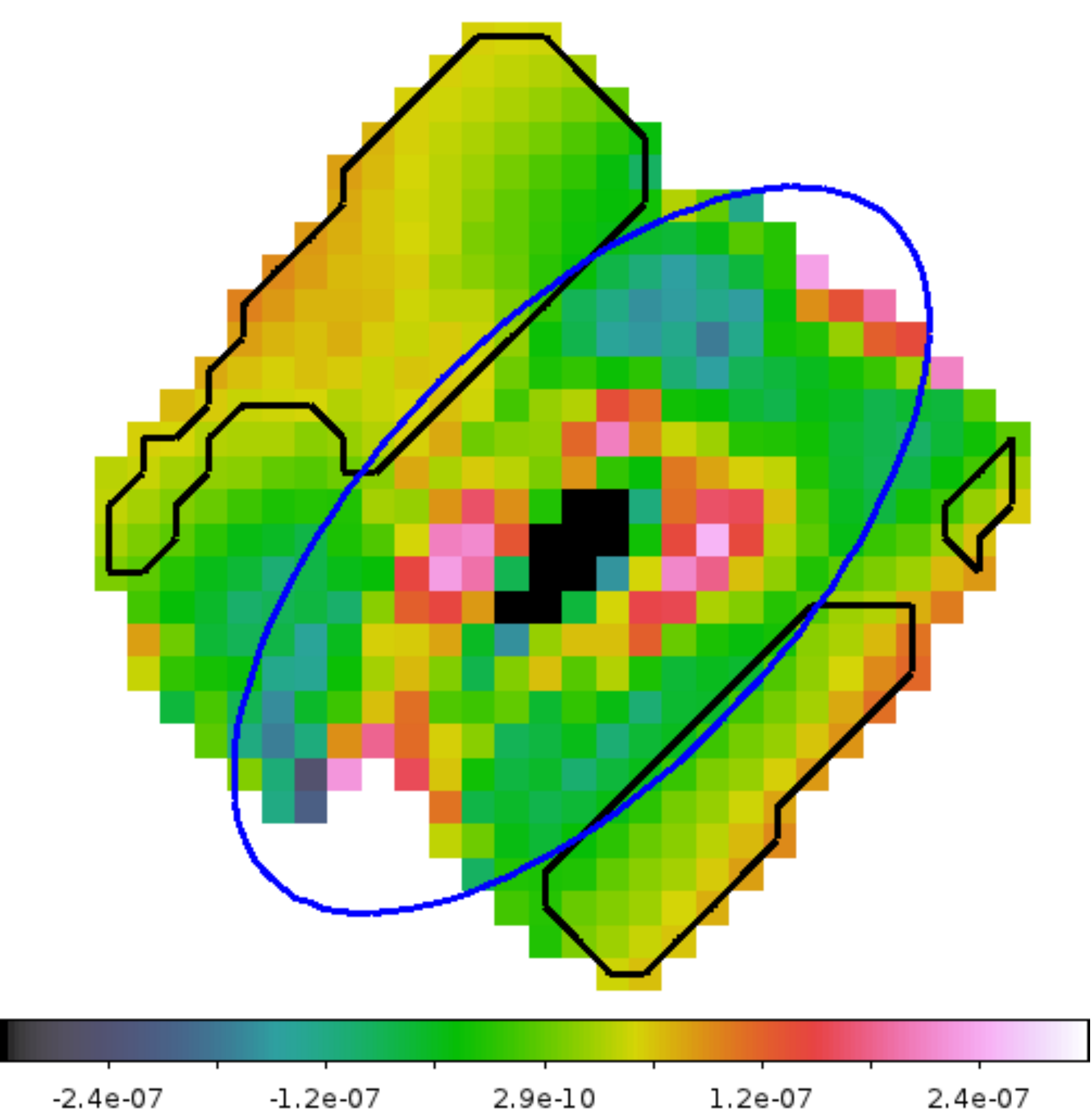}
\caption{To identify the amount of [CII] emission in the outflow attributable to the outflow itself, we model the [CII] intensity distribution (left) using only those regions with low velocity dispersion (outlined in black) with a two dimensional exponential disk (center).  The residual from this fit (right) show an excess in the regions above and below the disk (in blue) that correspond to the regions with increased velocity dispersion (in black).  From this we calculate an atomic gas outflow mass of 7.4 $\times 10^8$ M$_\sun$.  If we assume that all [CII] emission in regions of high velocity dispersion is due to the outflow (with no disk component), we calculate an atomic gas mass of $3.9 \times 10^9$ M$_\sun$. 
\label{fig:CII_outflow_fit}}
\end{figure*}

The high excitation is fully consistent with the effects of shocks, though we cannot rule out a contribution from other low level ionization sources.  We model this region using ITERA \citep{Groves2010} and the MAPPINGS III library of fast radiative shocks \citep{Allen2008}, and find it is well reproduced by relatively slow shock velocities between 100-200~\kms~(Figure \ref{fig:shocks}).    This modeled shock velocity is consistent between all three BPT diagrams, and agrees with previous examination of the shocks in this system \citep{Armus1995}.   Modeling is complicated by the combination of stellar photoionization within the disk with shocked region above the disk due to projections along the line of sight.  Given that the stellar photoionization contamination will move values down and to the left on the BPT diagram, thus resulting in lowered model shock velocities, we expect that the highest velocities measured $\sim$200~\kms~are most representative of the shock velocities throughout the superwind. 

Far-IR fine structure line ratios provide diagnostics for the physical conditions within photodissociation regions (e.g., \citealt{Kaufman1999}).  We map the diagnostic far-IR line ratios  [OI]/[CII], [OIII]/[CII], [OIII]/[OI] and [OIII]/[NII]  and, using the PACS 160~$\mu$m image as a proxy for the continuum, the [OI]/160~$\mu$m and [OIII]/160~$\mu$m ratios (Figure \ref{fig:PACSlineratios}).  Due to significant variations in the beam size with wavelength,  PACS line maps for each ratio are convolved to the lower resolution map in the pair using kernels from \cite{Aniano2011}.   The more excitation sensitive indicators, [OIII]/[CII] and [OIII]/[OI], show similar features, emphasizing the star formation dominated regions along the disk major axis.  The [OIII]/[CII] ratio shows a  value consistent with that found in the starburst ring of NGC 1097 \citep{Beirao2010} and with normal star forming galaxies \citep{Malhotra2001}. The morphology in the [OI]/[CII] map  is different from emission arising from the disk.  It follows the wind, suggesting that the [OI] may be enhanced in the wind, similar to the enhancement in the optical [OI] seen in the shocks.  This is further seen as [OI]/160~$\mu$m shows an increase that  aligns with the outflow region above the disk.  
[OIII]/160~$\mu$m is uniform except for an increase along the major axis at the field edge.  [OI]/[CII] shows much lower values (0.2-0.4) compared to M82 (\citealt{Contursi2013} ; see also Section \ref{sec:compM82}),  but is consistent with the value found by \cite{Appleton2013} in the shock-heated diffuse gas filament within Stephen's Quintet. 

\subsection{Mass entrainment in the outflow}
\label{sec:dust}
In M82 substantial amounts of dust and gas have been detected in the outflow.  We consider here the possibility of entrainment within the outflow of NGC~2146.

Assuming a \cite{Calzetti2000} attenuation law, \cite{Kreckel2013} converted the observed reddening in the Balmer decrement into a map of the V band attenuation, $A_{\rm V}$, across NGC~2146.  The Balmer line emission is able to probe down to $A_{\rm V}$ of 4-5~mag, corresponding to only a few percent of the Balmer line flux,  potentially only probing the surface of the dust layer and opaque to background emission.  However, given these limitations it appears that most of the dust lies in a relatively thin disk along the major axis but does exhibit a clumpy morphology  (see also Section \ref{sec:coout}).  

We trace the dust distribution through the far-IR continuum emission, detecting warm dust in the Spitzer MIPS 24~$\mu$m and SOFIA FORCAST 37~$\mu$m bands and the cold dust in the Herschel PACS 100~$\mu$m, 160~$\mu$m and SPIRE 250~$\mu$m bands (Figure \ref{fig:spire250}).    The emission is expected to be dominated by thermal emission from dust, and displays a disk-like morphology.    
The extent in the warm dust at 24~$\mu$m is well matched to the star formation as seen in the H$\alpha$ emission (Figure \ref{fig:opticallines}), including emission from outlying HII regions in the north.  The small amount of extended emission above and below the disk is a factor of five lower surface brightness than similar regions in M82 \citep{Engelbracht2006}, and unlike in M82 it displays a smooth morphology.  This region is also significantly contaminated by diffraction spikes due to the bright central source, and we do not consider this a reliable detection of entrained material.  The high resolution 37~$\mu$m emission traces the extent of the central starburst, with the region of peak emission unresolved at our 3\farcs5 resolution.  This is consistent with the unresolved central extent seen in the increased [OI] 63~$\mu$m velocity dispersion.    
In the cold dust we suffer from significantly lower angular resolution, however we generally see good agreement with the optical extent, with outlying features that generally align with the spiral arms.  Note that the hexagonal pattern at low flux limits in the 100 and 160~$\mu$m images is due to the shape of the PACS point spread function seen from the bright unresolved nuclear IR emission.  Careful modeling and subtraction of the disk emission would be necessary to distinguish emission due to dust in the wind (e.g. \citealt{Roussel2010}).  However from the morphology and lack of filamentary features we conclude that these observations present no strong evidence of substantial dust entrainment  at the size scales probed.

As a massive molecular outflow component has been previously identified in the superwind \citep{Tsai2009}, we estimate the atomic gas outflow mass in the superwind using the [CII] luminosity following \cite{Hailey-Dunsheath2010}  (see also \citealt{Maiolino2012}).   We estimate the mass as
\begin{equation}
\begin{split}
\frac{M_{atomic}}{M_\sun} &= 0.77 \left(\frac{0.7 L_{[CII]}}{L_\sun}\right) \left(\frac{1.4 \times 10^{-4}}{X_{C^+}}\right) \\
& \times \frac{1+2 e^{-91K/T}+n_{crit}/n}{2 e^{-91K/T}}
\end{split}
\end{equation}
where X$_{C^+}$ is the C$^+$ abundance per hydrogen atom, T is the gas temperature, n is the gas density and $n_{\rm crit}$ is the critical density of the [C II] 158~$\mu$m transition (3 $\times$ 10$^3$ cm$^{-3}$).  We assume a C$^+$ abundance typical of Milky Way photodissociation regions (X$_{C^+}$ = 1.6 $\times$ 10$^{-4}$; \citealt{Sofia2004}), 
a temperature of 200 K, and a density much higher than the critical density ($n_{crit}/n << 1$). We assume the high density case as this provides a lower limit for the total atomic gas mass.  
This calculation is relatively insensitive to the temperature chosen; a choice of 1000 K instead of 200 K would only result in a 15\% change in the total mass.   \cite{Sofia2011}, using a different methodology, suggests a lower value of X$_{C^+}$, which would also result in a more massive wind.  We emphasize that, in general, this mass estimate provides a lower limit in the case that we have a lower gas density or temperature.   

If we assume that all the [CII] emission from where we identify the superwind through the increased [CII] velocity dispersion is due to a massive outflow (with no disk component), we find L$_{[CII]} = 3.0 \times 10^9$ L$_\sun$  in the north-east and L$_{[CII]} = 1.5 \times 10^9$ L$_\sun$ in the south-west to give a total M$_{atomic} > 3.9 \times 10^9$ M$_\sun$.  However, given the 63$^\circ$ inclination, we may not be cleanly separating the disk contribution from the [CII] emission in the outflow.

As we cannot kinematically isolate the contribution of [CII]  emission due to the outflow from [CII] in a disk component, we fit the disk intensity in regions with low velocity dispersion using a two dimensional exponential disk model, and consider the residual emission in the regions identified as dominated by the superwind through the increased [CII] velocity dispersion (Figure \ref{fig:CII_outflow_fit}).  The residuals within the disk center are clumpy, however outside they are  smooth and indicate an excess of flux above the simple disk model.  
We note, however, that our disk modeling is sensitive to the high inclination angle of the disk, as in projection the 30\arcsec~distance angular extent observed along the minor axis covers a large radius ($\sim$6 kpc), and requires significant extrapolation from the central 5 kpc diameter region we are modeling.  
We find that that $\sim$20\% of the line flux is contributed by the outflow, which corresponds to L$_{[CII]} = 6.8 \pm 0.8 \times 10^8$~L$_\sun$ in the north-east and L$_{[CII]} = 1.9 \pm 0.3 \times 10^8$~L$_\sun$ in the south-west.  This gives a total atomic mass outflow of M$_{atomic} > 7.4 \times 10^8$ M$_\sun$, 40\% higher than what was found in the molecular superbubbles and outflows ($5.1 \times 10^8$ M$_\sun$;\citealt{Tsai2009}).  This is also a substantial amount as compared to the molecular ($2 \times 10^9$ M$_\sun$; \citealt{Young1988}) or the atomic hydrogen ($1.6 \times 10^9$ M$_\sun$; \citealt{Taramopoulos2001}) gas masses measured in the central disk.

\begin{figure*}[ht!]
\centering
\includegraphics[width=6in]{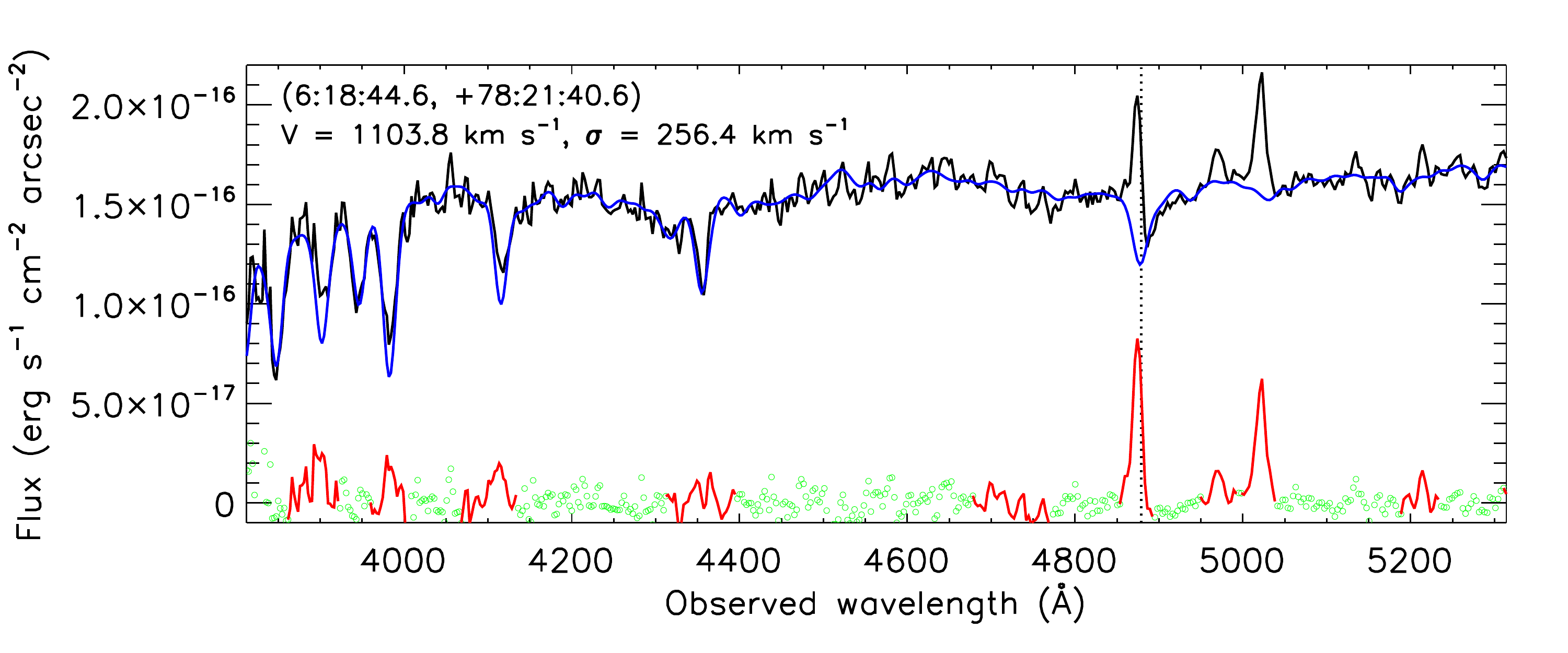}
\caption{The decoupling of the stellar and gas kinematics can be seen directly at the H$\beta$ 4861\AA ~feature (vertical dotted line) for one binned region (see Figure \ref{fig:optkinematics}) extracted from the optical IFU data cube.  Here we show the data (black) with the best fit stellar template (blue) and the residuals below (green).  Masked emission line features are shown in red.  The stellar absorption features are well fit across the spectrum with 50~\kms uncertainty, and the H$\beta$ line emission is clearly blueshifted from the underlying stellar absorption feature.   
\label{fig:skin}}
\end{figure*}

\begin{figure*}[ht!]
\centering
\includegraphics[width=7in]{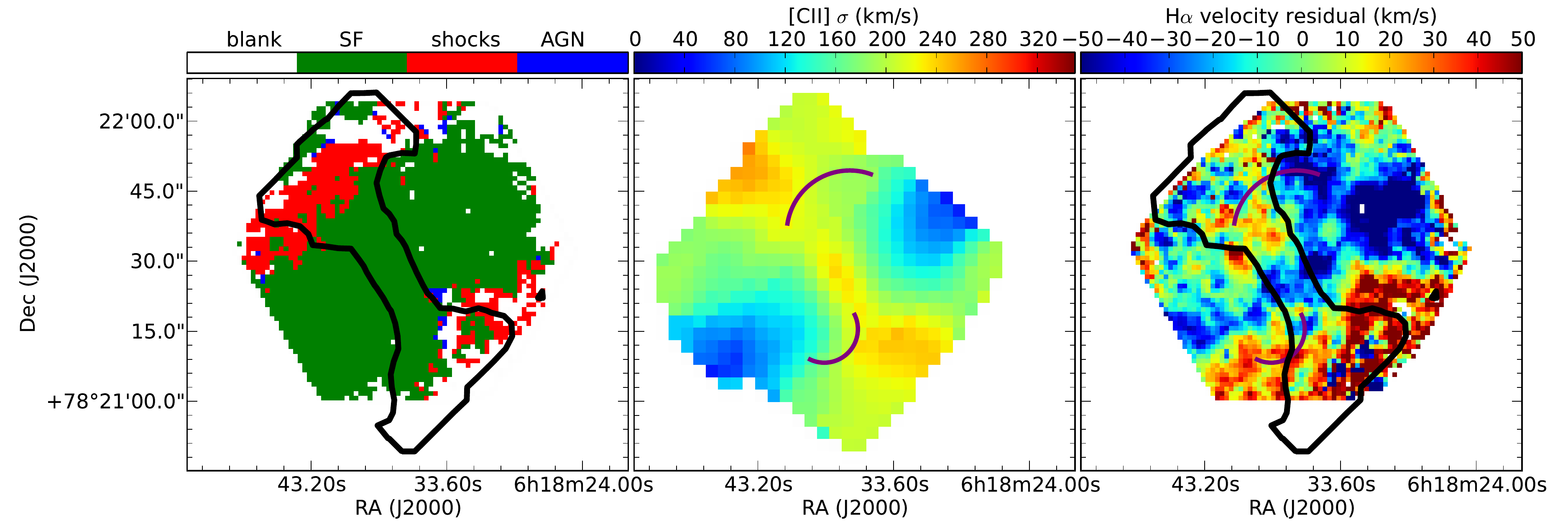}
\caption{A comparison of different superwind indicators.  Left: The optically identified shock regions (red, see Figure \ref{fig:bpt})  show very good agreement  with the high velocity dispersion identified in the [CII]  line (black contours).  Center:  The location (shown here with purple contours) of the superbubble  (south) and molecular outflow (north) identified by \cite{Tsai2009} appear offset from  the superwind identified by the increase in  [CII] velocity dispersion (in color).  Right: Comparing the location of the  H$\alpha$ velocity residual with the  superbubble and outflow identified in \cite{Tsai2009} (purple contours) and with the regions of increased CII velocity dispersion (black contours) reveals no alignment.  
\label{fig:overlay}}
\end{figure*}

\subsection{Stellar kinematics}
Our fits to the stellar kinematics show no ordered motions on 5~kpc scales in the central stellar disk within the $\sim$50~\kms~uncertainty, in stark contrast to the gas kinematics that show a large $\sim$500~\kms~velocity gradient across the same region.  Voronoi binning \citep{Cappellari2003} of neighboring spectra to ensure a minimum signal to noise of 20 in the stellar continuum confirms that this is not a signal to noise issue (Figure \ref{fig:optkinematics}, right).  However, we cannot exclude the presence of ordered motion on larger scales, or along the disk midplane where the dust extinction is extreme.  We directly observe the offset between the stellar and gas kinematics in individual binned regions by examining the H$\beta$ $\lambda$4861\AA~ feature (Figure \ref{fig:skin}).  Here, the underlying stellar absorption is clearly misaligned from the superposed nebular emission line.

This is in contrast to findings by previous studies in NGC~2146 of the stellar absorption features in longslit and driftscan spectra \citep{Kobulnicky2000, Greve2006}.  This may be due to significant dust absorption of the stellar continuum resulting in unreliable stellar kinematic measurements directly along the major axis.  We find that by simulating a longslit within our IFU data we can reproduce a rotation curve similar to that seen by \cite{Kobulnicky2000}, mainly by including the redshift emission to the north-west side of the disk, that covers a similar range of velocities but centered at a higher systemic velocity.  However, in the two dimensional kinematics it is apparent that this does not indicate bulk rotation in the stars, and that most of the motion appears unordered.  We cannot explain the $\sim$200~\kms~offset in systemic velocity that we observe (also apparent in the longslit measurements by \citealt{Kobulnicky2000}), particularly as we find it in relation to the line nebular velocities measured within our same dataset.  The overall agreement between the ionized gas kinematics observed in the far-IR, which should suffer very little dust attenuation, and in the optical (see Section \ref{sec:residvelfield} and Figure \ref{fig:resid}) suggests that any bias due to dust must be mainly affecting the stellar kinematics.  

\section{Discussion}
\label{sec:discussion}

\subsection{Comparing superwind indicators}
\label{sec:coout}

The case for a superwind in NGC~2146 was first made by \cite{Armus1995}, who identified soft X-ray emission consistent with swept up gas that has been shock heated.  They  also measured optical line diagnostics along the minor axis consistent with shock excitation and calculated that the energy from the starburst was enough to power the observed wind.  This soft X-ray emission has been further resolved \citep{dellaCeca1999, Inui2005}, and found to align with the CO bubbles identified by \cite{Tsai2009}.  Given our spatially resolved superwind indicators, we can compare these regions with the features we observe.  A clear diagram of the system geometry can be found in Figure 11 of \cite{Tsai2009}.

The regions consistent with shock contributions to the gas excitation (Figure \ref{fig:bpt}) align with the diffuse soft X-ray emission mapped by Chandra \citep{Inui2005}.  These regions are also relatively well aligned with the regions of increased velocity dispersion seen in the PACS line maps (Figure \ref{fig:overlay}, left).  This is fully consistent with a conical outflow both above and below the disk.  

The shocked region extends beyond the boundaries of the CO bubbles identified by \cite{Tsai2009}, and the bubbles appear misaligned with the regions of increased [CII] velocity dispersion (Figure \ref{fig:overlay}, center). Given the three dimensional nature of this outflow, the shock emission may lie along different filamentary channels than the molecular gas.  

The features in the H$\alpha$ residual velocity field seen above and below the disk after subtracting the  \cite{Greve2006} model rotation curve are very difficult to understand and do not align perfectly with the  outflow regions identified, as we would expect if they originated from the cone walls (Figure \ref{fig:overlay}, right).  The blueshifted region north of the disk falls mostly to the west of the outflow, and exists over a very narrow patch of that region.  The redshifted emission to the south-west does overlap  with the outflow, but is also offset to the west.  Perhaps these residual features are simply due to irregular motions within the clearly disturbed gas disk.  In addition, H$\alpha$ is quite susceptible to extinction, which will also influence the residuals measured.    We see no clear residuals in the PACS lines that would correspond to blueshifted or redshifted emission along the minor axis, however this is consistent with the expected geometry of the system (see Section \ref{sec:veldisp}).

\subsection{Comparison with M82}
\label{sec:compM82}
We find that NGC~2146 exhibits many  features similar to the prototype starburst galaxy M82.  Both benefit from a nearly edge-on viewing angle.  M82 also shows similar features through the soft X-ray emission tracing the starburst-driven wind \citep{Watson1984} and mass loaded outflows in the molecular gas \citep{Walter2002}.   Both systems display multiple line components in the ionized gas, extensively studied in M82 \citep{McKeith1995, Westmoquette2009a}.  Though in NGC~2146 they are not resolved in our H$\alpha$ data, they are apparent in the GHASP H$\alpha$ data cube \citep{Epinat2008}.  

In a comparison of the dimensions of the starburst and outflow between the two systems by \cite{Greve2000}, both display a similar geometry with conical outflows launched from a distributed region on the disk midplane and opening out above and below the disk.   NGC~2146 is a more massive galaxy, with a stellar mass of $2 \times 10^{10}$~M$_\sun$ \citep{Skibba2011} compared to $6 \times 10^9$~M$_\sun$ for M82 \citep{Karachentsev2004}.  Correspondingly, the outflow in NGC~2146 displays a significantly larger geometric scale in the dimensions of both the starburst and outflow regions \citep{Greve2000}, which serves to dilute the strength of the outflow through the larger starburst volume.  Our observations support this idea, as we measure in NGC~2146 a large opening angle of 90$^\circ$ and 120$^\circ$, compared to 50$^\circ$ for M82 \citep{Walter2002, Contursi2013}, and find no high velocity outflows in the H$\alpha$ residual velocity field for the inner $\sim$5 kpc region mapped.   

The gas excitation in M82 is dominated by stellar ionization close to the disk with an increasing contribution from shocks further away \citep{Heckman1990, Shopbell1998}.  We see this also very clearly in NGC~2146 (Figures \ref{fig:bpt} and \ref{fig:shocks}), as beyond 1 kpc above the disk the excitation is no longer consistent with photoionization and suggests a significant contribution from shocks.  

The far-IR fine structure lines in M82 were observed by  \cite{Contursi2013}.  In NGC~2146 we see that the [CII] and [OI] show quite clearly a $\sim$100~\kms~increase in the velocity dispersion in the superwind and along the minor axis to the disk center, very similar to what is seen in M82 and suggestive of moderate superwind velocities in the atomic gas.   Both galaxies also show a relatively wide opening angle in the [CII] and [OI] diagnostics.  In M82 this is different from what is seen in the H$\alpha$ outflow, which is much more collimated.  This is difficult to compare with NGC~2146, where we do not see the outflow clearly in H$\alpha$ in our observations of the central region. 

In M82 dust is seen up to 6 kpc along the minor axis both in mid-IR continuum and PAH emission \citep{Engelbracht2006}, although the physical properties modeled from the far-IR results suggest that much of it has been stripped by tidal interaction and not through the starburst wind \citep{Roussel2010}. The 24$\mu$m emission  in the wind of NGC~2146 is at least a factor of five lower surface brightness than M82 with a smoother morphology, providing no clear evidence for significant dust entrainment  (Figure \ref{fig:spire250}).

\subsection{Application to high redshift studies}
\label{sec:highz}

NGC~2146 is undergoing a substantial starburst, which places it as an outlier in specific SFR ( SFR / M$_*$  = $4 \times 10^{-10}$ year$^{-1}$) as a function of stellar mass when compared to local galaxies  \citep{Brinchmann2004}.  However, it is consistent with the increased specific SFR observed in high redshift z$\sim$1-3 galaxies \citep{Bauer2005} where outflows are commonly identified in starburst galaxies \citep{Pettini2001, Shapley2003, Steidel2004, Weiner2009, Rubin2010}.  
Observations of X-rays and ionized gas kinematics have proven to be strong indicators for the presence of outflows in galaxies in the local universe, however their use at high redshift is limited.  The [CII] line, a dominant coolant of the diffuse ISM that is observable with ALMA, provides a complimentary picture, tracing the morphology of the outflow through the widths of observed line profiles (see Section \ref{sec:veldisp}).

\begin{figure}[t]
\centering
\includegraphics[width=3.3in]{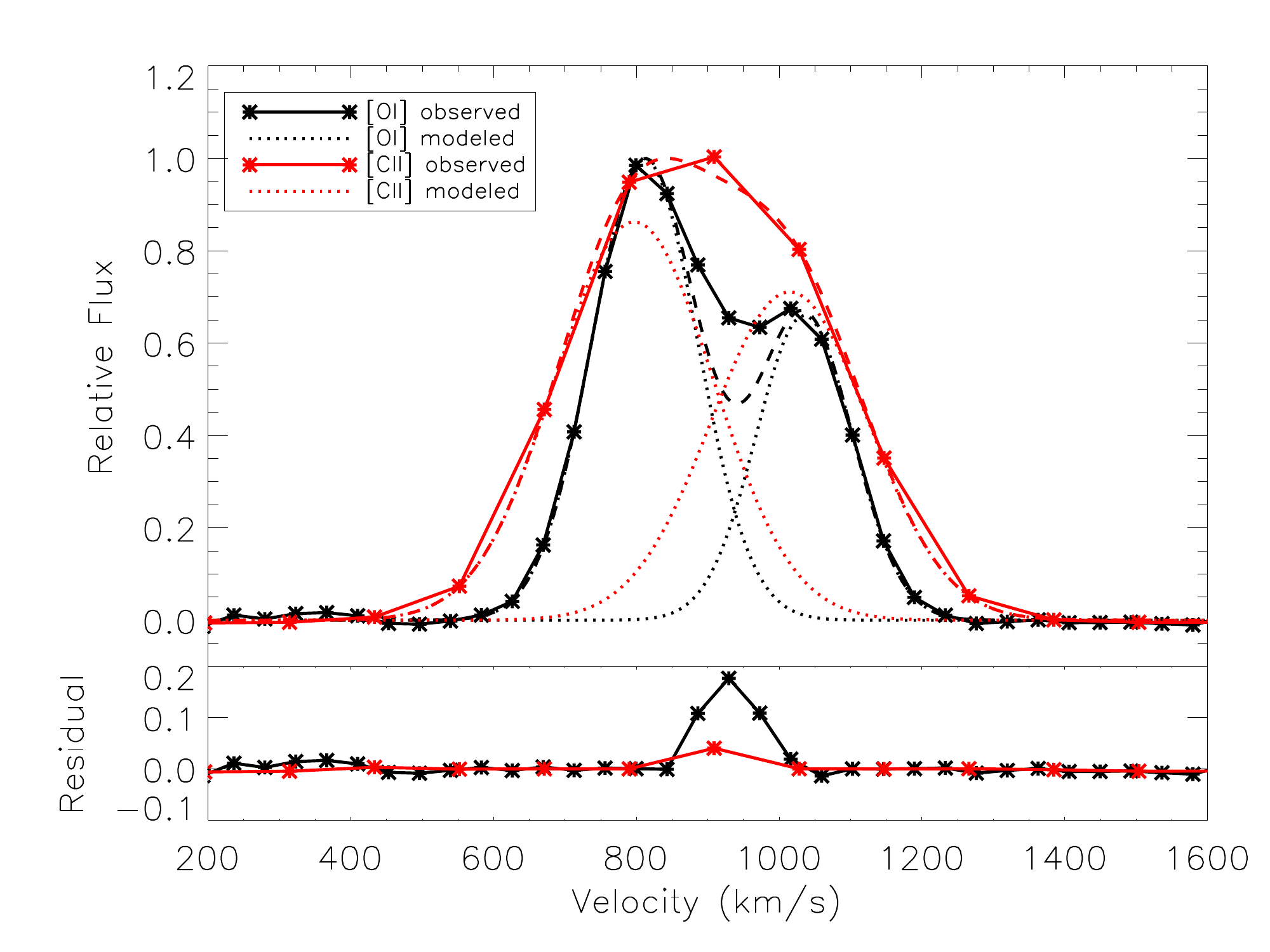}
\caption{Modeling of spatially integrated [OI] and [CII] line profiles assuming a double gaussian fit (top).  For [CII] and [OI] the fits recover the instrumental resolution for the approaching and receding sides of the disk, with no broad residual (bottom) that could indicate the contribution of the outflow.   
\label{fig:PACSopacity}}
\end{figure}

While for some systems an outflow may be identified from broad wings in the spectrum alone \citep{Maiolino2012, Carilli2013}, for high redshift galaxies similar to NGC~2146 spatial resolution is essential to identifying the outflow through the [CII] line emission.  In the spatially integrated far-IR line profile of NGC~2146 we find that both the [OI] and the [CII] lines are well fit by two Gaussians that decompose the galaxy into two roughly equal components that represent the motion of the disk and have widths consistent with the instrumental resolution  (Figure \ref{fig:PACSopacity}).    We recover no broad residual component that could be attributed to an outflow.  In general, outflows with a velocity equal to the circular velocity will be impossible to distinguish in unresolved data, which is especially problematic as in star formation dominated ULIRGs this is typically observed to be the case \citep{Martin2005}.  Spatially resolved observations are essential for identifying the presence of an outflow.

ALMA will be the perfect telescope for identifying superwinds through the [CII] line kinematics in high redshift systems, for which the galaxies M82 and NGC~2146 provide key local analogs.    The ALMA Band-9 and Band-10 receivers provide 0\farcs1-0\farcs2 resolution for the [CII] line at z=1-3, with a resulting 1-2 kpc spatial resolution that is equivalent to what we obtain here for NGC~2146. 

\subsection{A transitional galaxy}
\label{sec:elliptical}
The kinematic decoupling of the stars and gas are consistent with evidence from the optical morphology that this galaxy has undergone a recent merger event that may have disrupted the rotational support of the stars.   Disjoint optical morphology and kinematics have been tied in simulations to mergers with mass ratio between 4:1 and 10:1 \citep{Bournaud2004}.    This galaxy may be in the process of settling into an elliptical type galaxy, as is thought to happen with ULIRG galaxies \citep{Kormendy1992, Genzel2001}.

The interaction history of NGC~2146, however, is not completely clear.  There is no double-nucleus observed in the optical or IR images,  implying that the nuclei may have already merged, which could explain the odd stellar kinematics in the center.  Based on the extended H~\textsc{i} tail \citep{Taramopoulos2001}, it is possible that a third galaxy also passed by the system and was tidally disrupted, creating both the extended tidal features and possibly also causing features at smaller galactocentric radius such as the non-planar H$\alpha$ arc \citep{Greve2006}.  
In general, it is clear that NGC~2146 has not yet settled into a stable configuration, raising the question of what the final morphology of this galaxy will be.  We consider two possibilities: the substantial gas reservoir could serve to stabilize and rebuild the disk, resulting in a bulge-dominated spiral galaxy, or the starburst and wind could use up and expel sufficient material to stop star formation and convert the system to an elliptical.  

NGC~2146 has a substantial reservoir of gas in both the atomic and molecular phases surrounding the galaxy.  \cite{Young1988} measure a total mass of $1.2 \times 10^{10}$ M$_\sun$ in molecular clouds around the system, and \cite{Taramopoulos2001} detect a further $6.2 \times 10^9$ \Msun~in H~\textsc{i} in an extended configuration around the galaxy.  This mass in gas is roughly equal to the amount currently existing in the stars (M$_*$ = $2 \times 10^{10}$ \Msun; \citealt{Kennicutt2011}).  Even considering the current high SFR of 7.9~\Myr \citep{Kennicutt2011}, this galaxy has enough fuel to form stars for at least a further 2.3 Gyr.  Presumably it will form stars for much longer, assuming a lower SFR after the starburst ends, and will rebuild a disk of significant mass.

However, this system also has a substantial wind.  \Citet{dellaCeca1999} studied the soft X-ray emission from the starburst-driven superwind and estimated a mass loaded outflow rate of 9 \Myr, with energetics that suggest the hot gas may be able to escape from the galaxy.  Mass loading of the outflow is also observed by \cite{Tsai2009} in the molecular outflow and two superbubbles identified from the  molecular gas kinematics, and we estimate a substantial atomic gas mass of $7.4 \times 10^8$~M$_\sun$ in the outflow as well (see Section \ref{sec:dust}). If we consider that the starburst only affects the central gas, measured as $2 \times 10^9$ \Msun~ in molecular gas \citep{Young1988} and $1.6 \times 10^9$ \Msun~ in H~\textsc{i} \citep{Taramopoulos2001}, then the galaxy could potentially convert or expel all of this material in only 200 Myr, a reasonable duration for a starburst \citep{diMatteo2008}.  Even without removing all of the material, if the starburst is able to create a significant halo of hot gas it may be sufficient to shock heat any of the extended gas that falls in at later times.  The massive stellar spheroid may also provide the means to stabilize the gas against future star formation, through morphological quenching \citep{Martig2009}.

These two proposed scenarios are in fact extremes, with some combination of both likely responsible for the variety in the morphology and kinematics of elliptical galaxies that we see today \citep{Emsellem2011}.  Detailed simulations are necessary to distinguish what evolutionary future will play out in NGC~2146, but in either case it represents a very interesting and unusual transition object.

\section{Conclusion}
\label{sec:conclusion}

We present Herschel PACS far-IR fine structure line observations, optical IFU data and SOFIA 37 $\mu$m observations of the starburst driven superwind in NGC~2146, a nearby LIRG.  We find high velocity dispersions in all far-IR lines, with deconvolved linewidths of $\sim$250~\kms~that extend along the minor axis through the disk center and open into conical regions above and below the galaxy disk.  
This is fully consistent with the previously proposed picture of the outflow geometry that results in an alignment of the far side of the cone to the north and the near side of the cone to the south perpendicular to the line of sight \citep{Greve2000, Tsai2009}, with the line emission from the opposing cone walls overwhelmed in projection by the bright central disk.  
Unlike in optical studies, which are limited by the high extinction at the galaxy center, we are able to observe the superwind launching region, which extends less than kpc along the major axis in the center of the galaxy.

We present evidence for enhanced low-ionization line emission ([OI]/H$\alpha$, [SII]/H$\alpha$, [NII]/H$\alpha$) coincident with the regions of increased velocity dispersion above and below the plane seen in the far-IR emission line maps.  The optical line ratios are consistent with shock excitation, though we cannot rule out a contribution from other low level ionization sources.  The position of these shock and superwind indicators seen in the far-IR and optical have positions that match soft X-ray emission \citep{Inui2005}, with both the X-ray emission and CO bubbles \citep{Tsai2009} located within the full extent of the outflow cone we identify.  We do not detect dust entrainment in the outflow, in either the warm or the cold dust components.  In general, the outflow geometry is very similar to that observed in M82.  Outflows of this nature would not be easily discerned in unresolved [CII] observations of high redshift systems but will be resolved by ALMA at z$\sim$1-3.

We also observe stellar kinematics that are decoupled from the gas kinematics in all phases, consistent with predictions for post-merger systems, though we may also be suffering from some bias due to dust.    As the galaxy is still in the midst of its post-merger starburst it has yet to expel or transform the bulk of its molecular gas.  The starburst driven wind is potentially crucial for transformation of the galaxy's morphology into a red and dead elliptical.

Far-IR line observations provide an unobscured view of the outflow and starburst, crucial for detailed study of the physical conditions in these regions.  This system will provide an important local analog for future high redshift studies of IR-luminous galaxies  driving galactic-scale superwinds. 

\acknowledgements

We would like to thank the referee for their helpful comments.  KK acknowledges the support of grants GR 3948/1-1 and SCHI 536/8-1 from the DFG Priority Program 1573, ``The Physics of the Interstellar Medium".  
ADB acknowledges support from the National Science Foundation through grant AST-0955836, as well as a Cottrell Scholar award from the Research Corporation for Science Advancement.  

This work is based on observations made with Herschel.  Herschel is an ESA space observatory with science instruments provided by European-led Principal Investigator consortia and with important participation from NASA.
PACS has been developed by a consortium of institutes led by MPE (Germany) and including UVIE (Austria); KU Leuven, CSL, IMEC (Belgium); CEA, LAM (France); MPIA (Germany); INAF-IFSI/OAA/OAP/OAT, LENS, SISSA (Italy); IAC (Spain). This development has been supported by the funding agencies BMVIT (Austria), ESA-PRODEX (Belgium), CEA/CNES (France), DLR (Germany), ASI/INAF (Italy), and CICYT/MCYT (Spain).

Based in part on observations collected at the Centro Astron\'{o}mico Hispano Alem\'{a}n (CAHA), operated jointly by the Max-Planck Institut f\"{u}r Astronomie and the Instituto de Astrofisica de Andalucia (CSIC).  

Based in part on observations made with the NASA/DLR Stratospheric Observatory for Infrared Astronomy (SOFIA). SOFIA is jointly operated by the Universities Space Research Association, Inc. (USRA), under NASA contract NAS2-97001, and the Deutsches SOFIA Institut (DSI) under DLR contract 50 OK 0901 to the University of Stuttgart.  

This research made use of APLpy, an open-source plotting package for Python hosted at http://aplpy.github.com.  

\bibliographystyle{apj}

\end{document}